\documentclass[journal,twocolumn]{IEEEtran}
\usepackage{amsmath, amsfonts, amssymb}
\usepackage[T1]{fontenc}
\usepackage{graphicx}
\usepackage[dvipsnames]{xcolor}
\usepackage[utf8]{inputenc}
\usepackage{color}
\usepackage{multirow, ,bigstrut}
\usepackage{caption}
\usepackage{tikz}
\usetikzlibrary{fit}
\usepackage{pgfplots}
\pgfplotsset{compat=1.7}
\usepackage{pgfplotstable}
\usepackage{svg}
\usepackage[caption=false]{subfig}
\usepackage{xcolor,colortbl}
\usepackage{pgf-pie}
\usepackage[export]{adjustbox}
\usepackage{wrapfig}
\pdfoutput=1
\UseRawInputEncoding 
\usepackage{url}

\usepackage[breaklinks]{hyperref}
\usepackage{breakurl}
\usepackage{fancyhdr}
\usepackage{cite}
\usepackage{tikz,tikz-3dplot}
\usetikzlibrary{3d,decorations.text,shapes.arrows,positioning,fit,backgrounds, shapes}
\hypersetup{breaklinks=true}
\usetikzlibrary{calc}
\usepackage{mathptmx}
\usepackage{gnuplottex}
\usepackage{lmodern} 
\usepackage{tikz-3dplot}
\usepackage{lscape}
\usepackage[normalem]{ulem}
\definecolor{cerisepink}{rgb}{0.93, 0.23, 0.51}
\definecolor{color1}{RGB}{255, 253, 204}

\newsavebox\audiosignal
\savebox\audiosignal{%
\begin{tikzpicture}[
    anchor=center,
    declare function={
      excitation(\t,\w) = sin(\t*\w)*(1.5);
      noise = rnd - 0.5;
      source(\t) = excitation(\t,20) + noise;
      filter(\t) = 0.5 - abs(sin(mod(\t, 50)));
      speech(\t) = 1 + source(\t)*filter(\t);
    }
  ]
    \draw[help lines] (0,1) -- (3,1);
    \draw[blue, thick, x=0.0085cm, y=1cm] (0,1) -- plot [domain=0:360, samples=144, smooth] (\x,{speech(\x)});
\end{tikzpicture}%
}
\newsavebox\segaudiosignal
\savebox\segaudiosignal{%
\begin{tikzpicture}[
    anchor=center,
    declare function={
      excitation(\t,\w) = sin(\t*\w)*(1.5);
      noise = rnd - 0.5;
      source(\t) = excitation(\t,20) + noise;
      filter(\t) = 0.5 - abs(sin(mod(\t, 50)));
      speech(\t) = 1 + source(\t)*filter(\t);
    }
  ]
    \draw[help lines] (0,1) -- (3,1);
    \draw[blue, thick, x=0.0085cm, y=1cm] (0,1) -- plot [domain=0:90, samples=144, smooth] (\x,{speech(\x)});
    \draw[blue, thick, x=0.0085cm, y=1cm] (0,1) -- plot [domain=180:270, samples=144, smooth] (\x,{speech(\x)});
    \draw[black!10, thick, x=0.0085cm, y=1cm] (0,1) -- plot [domain=270:360, samples=144, smooth] (\x,1);
    \draw[red, ultra thick, x=0.0085cm, y=1cm] (0,1) -- (0,1.85)-- (90,1.85) --(90,1) -- (180,1);
    \draw[red, ultra thick, x=0.0085cm, y=1cm] (180,1) -- (180,1.85)-- (270,1.85) --(270,1) -- (360,1);
\end{tikzpicture}
}
\title{A Generic Deep Learning Based Cough Analysis System from Clinically Validated Samples for Point-of-Need Covid-19 Test and Severity Levels}
\author{Javier Andreu-Perez,~\IEEEmembership{SMIEEE}, Humberto P\'erez-Espinosa, Eva Timonet, Mehrin Kiani, Manuel I. Gir\'on-P\'erez, Alma B. Benitez-Trinidad, Delaram Jarchi, \emph{SMIEEE}, Alejandro Rosales-P\'erez, Nick Gatzoulis, Orion F. Reyes-Galaviz,  Alejandro Torres-Garc\'ia, Carlos A. Reyes-Garc\'ia, Zulfiqar Ali, Francisco Rivas}

\date{May 2020}
\newcommand{\FiguresDir}{./Figures}

\begin{document}
\bstctlcite{IEEEexample:BSTcontrol}

\maketitle
\thispagestyle{fancy}
\begin{abstract}
In an attempt to reduce the infection rate of the COrona VIrus Disease-19 (Covid-19) countries around the world have echoed the exigency for an economical, accessible, point-of-need diagnostic test to identify Covid-19 carriers so that they (individuals who test positive) can be advised to self isolate rather than the entire community. Availability of a quick turn-around time diagnostic test would essentially mean that life, in general, can return to normality-at-large. In this regards, studies concurrent in time with ours have investigated different respiratory sounds, including cough, to recognise potential Covid-19 carriers. However, these studies lack clinical control and rely on Internet users confirming their test results in a web questionnaire (crowdsourcing) thus rendering their analysis inadequate.
We seek to evaluate the detection performance of a primary screening tool of Covid-19 solely based on the cough sound from \emph{8,380} \emph{clinically validated samples with laboratory molecular-test} (\emph{2,339} Covid-19 positive and \emph{6,041} Covid-19 negative) under quantitative RT-PCR (qRT-PCR) from certified laboratories. All collected samples were clinically labelled, i.e. Covid-19 positive or negative, according to the results in addition to the disease severity based on the qRT-PCR threshold cycle (Ct) and lymphocytes count from the patients. Our proposed generic method is an algorithm based on Empirical Mode Decomposition (EMD) for cough sound detection with subsequent classification based on a tensor of audio sonographs and deep artificial neural network classifier with convolutional layers called \emph{`DeepCough'}. Two different versions of DeepCough based on the number of tensor dimensions, i.e. DeepCough2D and DeepCough3D, have been investigated. These methods have been deployed in a multi-platform prototype web-app \emph{`CoughDetect'}. 
Covid-19 recognition results rates achieved a promising AUC (Area Under Curve) of $98.80\%\pm 0.83\%$, sensitivity of $96.43\%\pm 1.85\%$, and specificity of $96.20\%\pm 1.74\%$ and average AUC of $81.08\%\pm 5.05\%$ for the recognition of three severity levels. 
Our proposed web tool as a point-of-need primary diagnostic test for Covid-19 facilitates the rapid detection of the infection. We believe it has the potential to significantly hamper the Covid-19 pandemic across the world.
\end{abstract}
\section{Introduction}
\label{section:intro}
The COrona VIrus Disease-19 (Covid-19) is an infectious disease caused by the newly discovered severe acute respiratory syndrome coronavirus 2 (SARS-CoV-2). Covid-19 bears stark similarities with the Severe Acute Respiratory Syndrome (SARS) as well as the common cold. According to the World Health Organization (WHO), the mild symptoms of Covid-19 can include fever, cough and shortness of breath akin to the common cold \cite{whocoronavirus}. Like SARS in more severe cases, Covid-19 also causes pneumonia and/or significant breathing difficulties, and in some rare instances, the disease can be fatal with the overall mortality rate estimated to be 0.28\% worldwide. The initial cases of Covid-19 were initially diagnosed as \emph{pneumonia} on 31 December 2019, and later re-diagnosed as Covid-19.

Covid-19 has proven to be a very infectious disease with the virus (SARS-CoV-2) spreading quickly on coming in close contact with an infected person (mean infection rate of 2.5). More specifically, according to the WHO, the (Covid-19) virus is transmitted through direct contact with respiratory droplets of an infected person (generated through coughing and sneezing) \cite{WHO_Covidspread}. The WHO declared it a global pandemic on 11 March 2020, within three months of first reported cases in China.

Covid-19 has put considerable strain on the health systems worldwide, with even developed countries struggling to test enough people to stop its spread effectively. Hence, taking the Covid-19 pandemic context in consideration, it is important to re-think the classical approaches for timely case finding \cite{bedford2020covid}, as well as to utilise the limited resources available most effectively \cite{CohenJ2020}.

In past pandemics, such as Malaria, a two pronged approach for a screening test was successfully employed to combat the spread of a prevalent virus \cite{VialH2013}. In these two-stage strategies, the primary stage focuses on greater accessibility and ease of screening that is cost-effective. The primary stage is to `alert' a potential carrier if they test positive on a primary screening test. In most cases, only those who test positive on the primary test go on to the secondary test, hence reducing the burden on the health system, and making the most of the resources available to conduct the secondary test.

The secondary screening is where the null hypothesis that the participant is not carrying an infection is accepted or rejected. The current techniques employed for screening of Covid-19 use serology, and diagnosis is based on the presence of genetic material of the virus. Clinical molecular tests have robust diagnostic accuracy but require specialised equipment, as well as trained personnel to conduct the test. The turn-around time of these tests can vary from hours to several days. 

Given the established success of two-pronged screening mechanisms to hamper the spread of infectious diseases, in this work, \emph{we aim to develop a web-based tool for the primary screening of Covid-19}. The motivation is to identify Covid-19 carriers using a model trained with clinically validated cough signals since Covid-19 affects the respiratory system \cite{Eliezer2020, li2020covid, latif2020leveraging}. Established works have evidenced the possibility of using the latent sound characteristics of coughs to identify respiratory diseases \cite{SwarnkarV2012, ChatrzarrinH2011}. In addition, prior works have also reported that voluntary coughs (asymptomatic) contain sound characteristics that allow detecting abnormal pulmonary functioning and respiratory diseases \cite{abaza2009,infante2017}.

The remainder of this paper is structured as follows: section \ref{section:contributions} outlines a summary of contributions, section \ref{section:relatedwork} gives an overview of related work; section \ref{section:methods} describes the procedural and methodological stages of the development of this technology; section \ref{section:results} evaluates the recognition and assessment results; section \ref{section:discussion} discusses the results and achievements; with conclusion in section \ref{section:conclusion}.

\section{Summary of contributions}
\label{section:contributions}
The main contributions of this work are manifold and listed as follows:
\begin{enumerate}
    \item The proposed method \emph{`DeepCough'} achieves high accuracy, without the necessity of using specific pre-trained models or transfer learning of data from other studies. Hence, differently from related work, \textbf{the proposed methodology is generic}, paving the way for derivative works.
    \item In contrast to related work, we are able to evaluate the real capacities of detecting Covid-19 in a large clinically validated dataset (8,000+) where \textbf{all data samples are matched with molecular-test of Covid-19 viral infection} dispensed in certified laboratories to participants.
    \item Also unique to this work, the accompanying molecular-tests (qRT-PCR) along with the cough samples, allow us \textbf{to predict as well the extent of the infection}. This is studied in this work using either the cycle threshold (Ct) from the qRT-PCR test or lymphocyte counts.
    \item Furthermore, a \textbf{full-stack automatic processing framework, from a raw sound stream to the test results,} is also presented.
    \item Development of \textbf{a tangible test service prototype}, as a platform-independent web-app service, \emph{CoughDetect.com}\footnote{https://coughdetect.com}
\end{enumerate}

\section{Related Work}
\label{section:relatedwork}
In an attempt to better understand the Covid-19 infection, and its associated symptoms, scientists have been collecting a wide spectrum of information in the latest months. This includes, but is not limited to, the respiratory sounds related to Covid-19 \cite{Eliezer2020} \cite{latif2020leveraging} \cite{Xu2020_pathological}, thermal imaging \cite{WeiTing2020}, digestive symptoms \cite{Pan2020}, as well as self reported surveys. The motivation of collating Covid-19 related information is to develop robust mechanisms for early detection of Covid-19. The most common symptoms of Covid-19 have been linked to pneumonia (cough, fever, shortness of breath, among others). Therefore, the analysis of cough audio signals is considered a viable course of action for a primary Covid-19 diagnosis \cite{latif2020leveraging}. 

In general, three different respiratory sounds have been investigated to detect Covid-19 in patients: voice, breath and cough. The voice is a bio-signal that has been studied for many years to decode emotional, mental and physical aspects of a speaker. Usman \emph{et al.} \cite{Usman2020} conclude that there is a strong correlation between speech and Covid-19 symptoms, and therefore endorse the usage of speech signals for detecting Covid-19.  

Faezipour \emph{et al.} \cite{faezipour2020smartphone} recommended the use of signal processing techniques in tandem with state-of-the-art machine learning and pattern recognition techniques for preliminary diagnosis of Covid-19 from breathing audio signals. However, neither of the studies \cite{Usman2020} and \cite{faezipour2020smartphone} encompass the Covid-19 recognition at this stage, with the additional caveat of quality of breath sounds hinged on the sensitivity of the microphone. 

Another notable work on breathing patterns is done by Wang \emph {et al.} \cite{wang2020abnormal} who developed a respiratory simulation model (RSM) for detecting the abnormal respiratory patterns of people remotely, and unobtrusively using a depth camera. However, their proposed RSM did not incorporate data from Covid-19 carriers. Nevertheless, the use of video cameras may raise privacy concerns. Imran et al. \cite{imran2020ai4covid} presented AI4COVID - an approach to classify coughs using deep learning, and achieved an accuracy of 92.85\%. However, their dataset contains only 70 Covid-19 cough samples, which renders their analysis to be inconclusive.

Sharma \emph{et al.} \cite{sharma2020coswara} presented \emph{Coswara}\footnote{https://coswara.iisc.ac.in/}, a  database embodying respiratory sounds (cough, breath, and voice). This dataset is crowdsourced (volunteers from the web), i.e. not clinically controlled samples, with only eight positive Covid-19 samples at the time of writing of this study. 
Here, it is also important to note that sound modalities, especially voice, embodies privacy concerns since an individual can be identified from their voice \cite{hollien2002forensic}.
Other notable database creation projects collecting data from the web include: Opensigma\footnote{https://opensigma.mit.edu/} by MIT collects collecting cough samples, Corona Voice Detect\footnote{https://voca.ai/corona-virus/} by Voca.ai and Carnegie Mellon University (CMU) is collecting voice data, Covid Voice Detector\footnote{https://cvd.lti.cmu.edu/} also by CMU is collecting further voice samples, and finally, the Covid-19 Sounds App\footnote{https://www.covid-19-sounds.org/} by the University of Cambridge is collecting crowdsourced samples of voice, cough, and breath.

A consensus derived from the related work referenced above is the challenge associated in the collection of clinically validated Covid-19 data which can be subsequently used for the training of Covid-19 recognition mechanisms. Towards this end, the data used in this study is collected following a strict protocol designed at laboratories and hospitals dedicated to Covid-19 diagnosis by expert immunologists. Another major strength of our proposed web-based app \emph{CoughDetect} lies in the anonymity of the users. Coughs sounds are inherently anonymous. Collecting just cough sounds, along with the usage of in-house code only and strict privacy-preserving practices, we have ensured that participants share their cough samples without exposing their personal information. This robust quality control of our collected samples is an advantage of our work with respect to other studies, e.g. collecting clinical data via web questionnaires (crowdsourcing).

\section{Methods for developing a point-of-need Covid-19 web-app service from only cough sound samples}
\label{section:methods}
\begin{figure}[!b]
  \centering
\subfloat[Mobile Phone]{
\fcolorbox{black}{black}{\includegraphics[scale = 0.25, clip, trim = 12cm 3cm 13cm 2.5cm ]{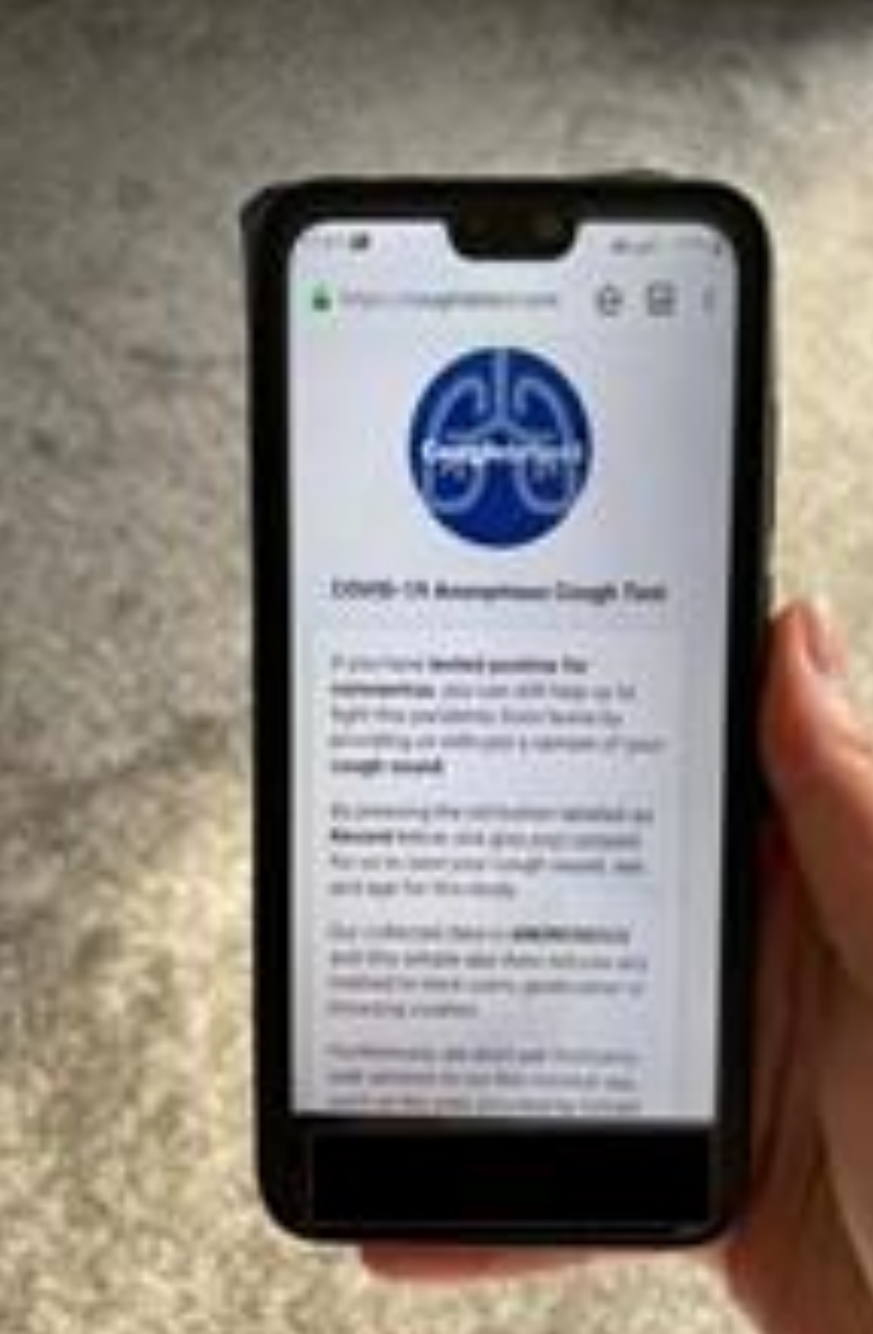}}
}%
\subfloat[Laptop]{
\fcolorbox{black}{black}{\includegraphics[scale = 0.25, clip, trim = 6cm 3cm 19cm 2.5cm]{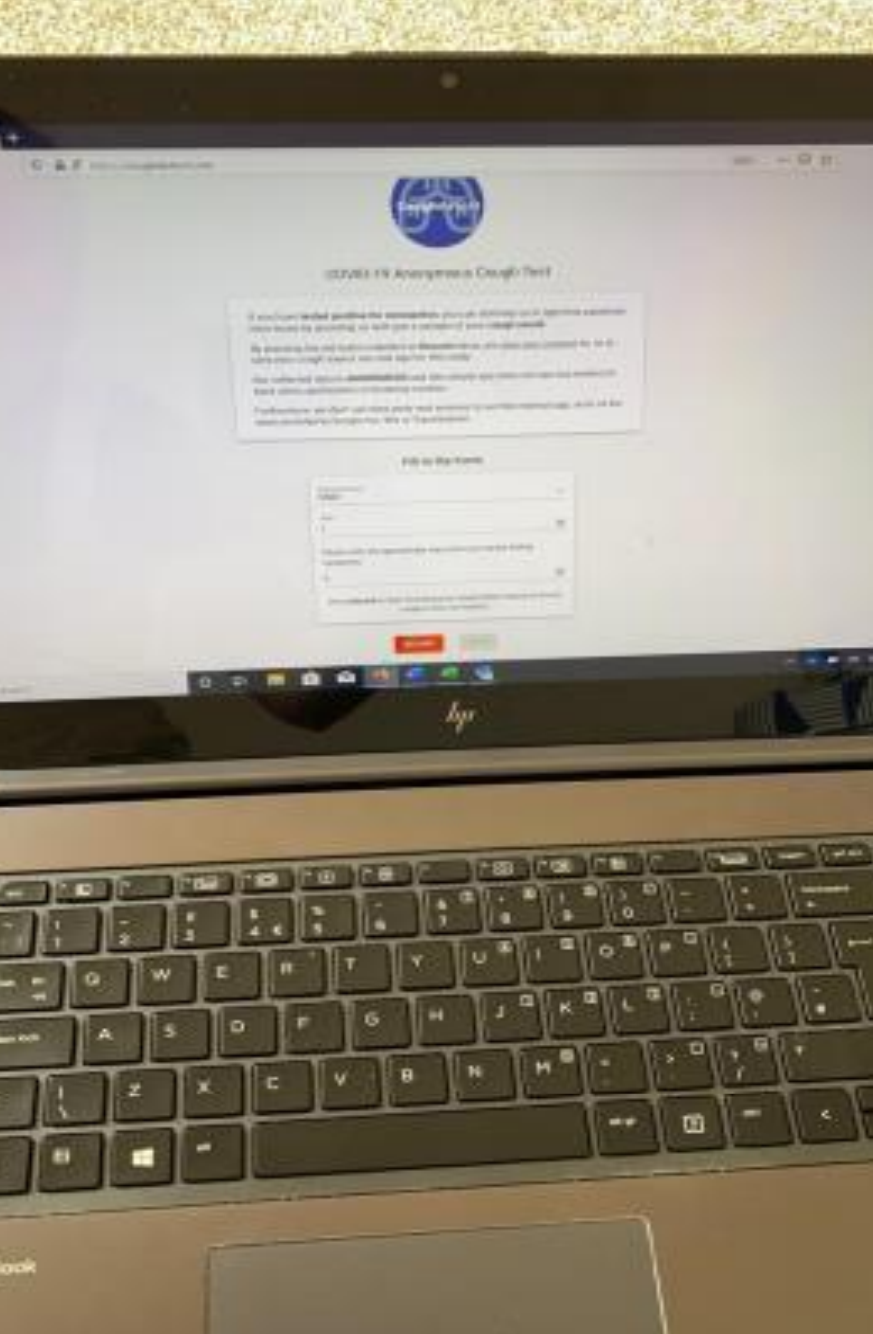}}
}%
\subfloat[Tablet]{
\fcolorbox{black}{black}{\includegraphics[scale = 0.25, clip, trim = 16cm 3cm 7cm 2.5cm]{Figures/App_LaptopandTablet.pdf}}
}%
\caption{The CoughDetect app can be easily used with a a) mobile phone, b) laptop or c) tablet connected with Internet.}
\label{fig:CoughDetectApp_Devices}
\end{figure}

The cough samples are collected by means of an in-house developed web app named \href{https://coughdetect.com}{CoughDetect}. The CoughDetect app (\url{https://coughdetect.com}) can be easily used with a laptop, mobile phone, or tablet, as shown in Fig. \ref{fig:CoughDetectApp_Devices}. The development of the whole stack for Covid-19 primary screening required the use of several technologies to capture, process, analyse and make the test available. An illustration of the proposed technology stack diagram for the CoughDetect operational architecture is shown in Fig. \ref{fig:CoughDetectApp}. The app records (\texttt{.wav}) sound files at \texttt{44,100Hz} sample rate and transfers them to a secure data server using \texttt{HTTP} over \texttt{SSL} connection. 

The three stages of the development stack include:
\begin{enumerate}
    \item Sound stream processing and Detection;
    \item A recognition method based on the generation of an Acoustic Cough tensor and Deep Learning (DeepCough);
    \item Development and Deployment of the framework in a Web Tool App (CoughDetect).
\end{enumerate}

A flow chart delineating the steps in the inference mechanism of DeepCough is shown in Fig. \ref{fig:COVIDflowchart}. The pre-processing of the raw sound signals is done to increase the signal-to-noise ratio and reduce the signal size. Cough bursts are detected in the recording and the rest of the signal is discarded. A set of low-level acoustic descriptors (a.k.a. sonographs) are extracted from a pre-processed cough sound. Two- and three-dimensional (2D and 3D) tensors are generated from these descriptors. These tensors are fed to a convolutional deep neural network that allows classification of positive and negative Covid-19 cough samples. Additionally, positive patients are sub-classified according to severity: borderline positive, standard positive, high positive based on qRT-PCR values and lymphopenia, or normal lymphocytes based on their blood lymphocyte count, as shown in Fig. \ref{fig:COVIDflowchart}. Further details of research ethics and the different stages for building the CoughTensor and classification are presented next.

\tikzstyle{decision} = [diamond, draw, fill=blue!20, 
    text width=4.5em, text badly centered, node distance=3cm, inner sep=0pt]
\tikzstyle{block} = [rectangle, draw,  
    text width=2.8cm, text centered, rounded corners, minimum height=1.5cm]
\tikzstyle{line} = [draw, -latex']
\tikzstyle{cloud} = [draw, ellipse,fill=red!20, node distance=3cm,
    minimum height=2em]
\tikzstyle{dotted_block} = [draw=black!30!white, line width=1pt, dash pattern=on 1pt off 4pt on 6pt off 4pt, inner ysep=1mm,inner xsep=1mm, rectangle, rounded corners ]
\tikzstyle{myarrows} = [draw=black,solid,line width=.5mm, ->]
\tikzstyle{dashedline} = [draw=black,dashed,line width=.5mm]

\begin{figure*}[!tbp]
\centering
\scalebox{0.5}{
\begin{tikzpicture}[node distance = .5cm, auto]
\node (FrontEnd_logo) [text width=3cm] {\includegraphics[ scale=0.25, clip, trim = 0.5cm 1cm 0.5cm 1.2cm ]{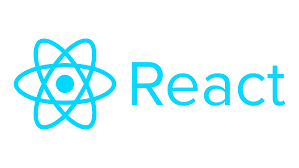}};
\node (FrontEnd_text) [text width=3cm, below of =FrontEnd_logo, node distance = 1.2cm ] {\texttt{\textbf{Front End\\} React App \\ Audio Recording}};
\node (FEnd) [dotted_block, inner ysep=2mm, inner xsep=2mm, fit= (FrontEnd_logo)(FrontEnd_text)] {};
\node (BackEnd_logo) [text width=3cm, right of =FrontEnd_logo, node distance = 6cm] {\includegraphics[ scale=0.2, clip, trim = 0.5cm .5cm 0.5cm 1.2cm ]{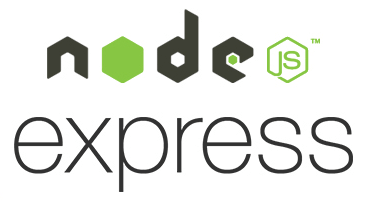}};
\node (BackEnd_text) [text width=3cm, below of =BackEnd_logo, node distance = 1.3cm ] { \texttt{\textbf{Back End\\}Node/Express \\ API Server}};
\node (BEnd) [dotted_block, inner ysep=2mm, inner xsep=2mm, fit= (BackEnd_logo)(BackEnd_text)] {};
\node (PythonScript_logo) [text width=3cm, right of =BackEnd_logo, node distance = 5cm, yshift = 1.5cm] {\includegraphics[ scale=0.05, clip, trim = 0.5cm .5cm 0.5cm 1.2cm ]{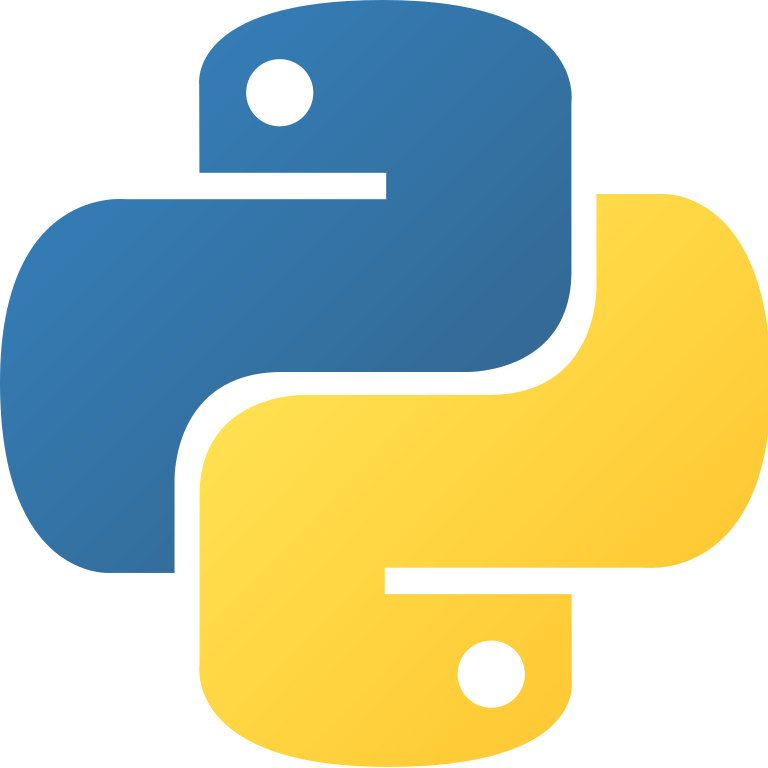}};
\node (PythonScript_text) [text width=3.3cm, right of =PythonScript_logo, node distance = 1.8cm ] {\texttt{\textbf{Python Script\\} Machine Learning \\ Signal Processing}};
\node (PS) [dotted_block, inner ysep=2mm, inner xsep=2mm, fit= (PythonScript_logo)(PythonScript_text)] {};
\node (Database_logo) [text width=4cm, below of =PS, node distance = 4.5cm] {\includegraphics[ scale=0.25, clip, trim = 1cm .5cm 1cm 3cm ]{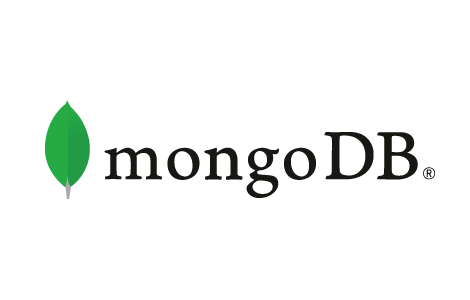}};
\node (Database_text) [text width=4.5cm, below of =Database_logo, node distance = .5cm ] {\texttt{\textbf{Database} MongoDB Instance}};
\node (MDB) [dotted_block, inner ysep=1mm, inner xsep=2mm, fit= (Database_logo)(Database_text)] {};
\node (App_Arch) [dotted_block, inner ysep=5mm, inner xsep=3mm, fit= (MDB)(PS)(BEnd)(FEnd)] {};
\node (App_screenshot) [block, node distance = 1.5cm ,text width=4cm, label= \large CoughDetect Web and Mobile App, right = of App_Arch] {\includegraphics[scale=0.2]{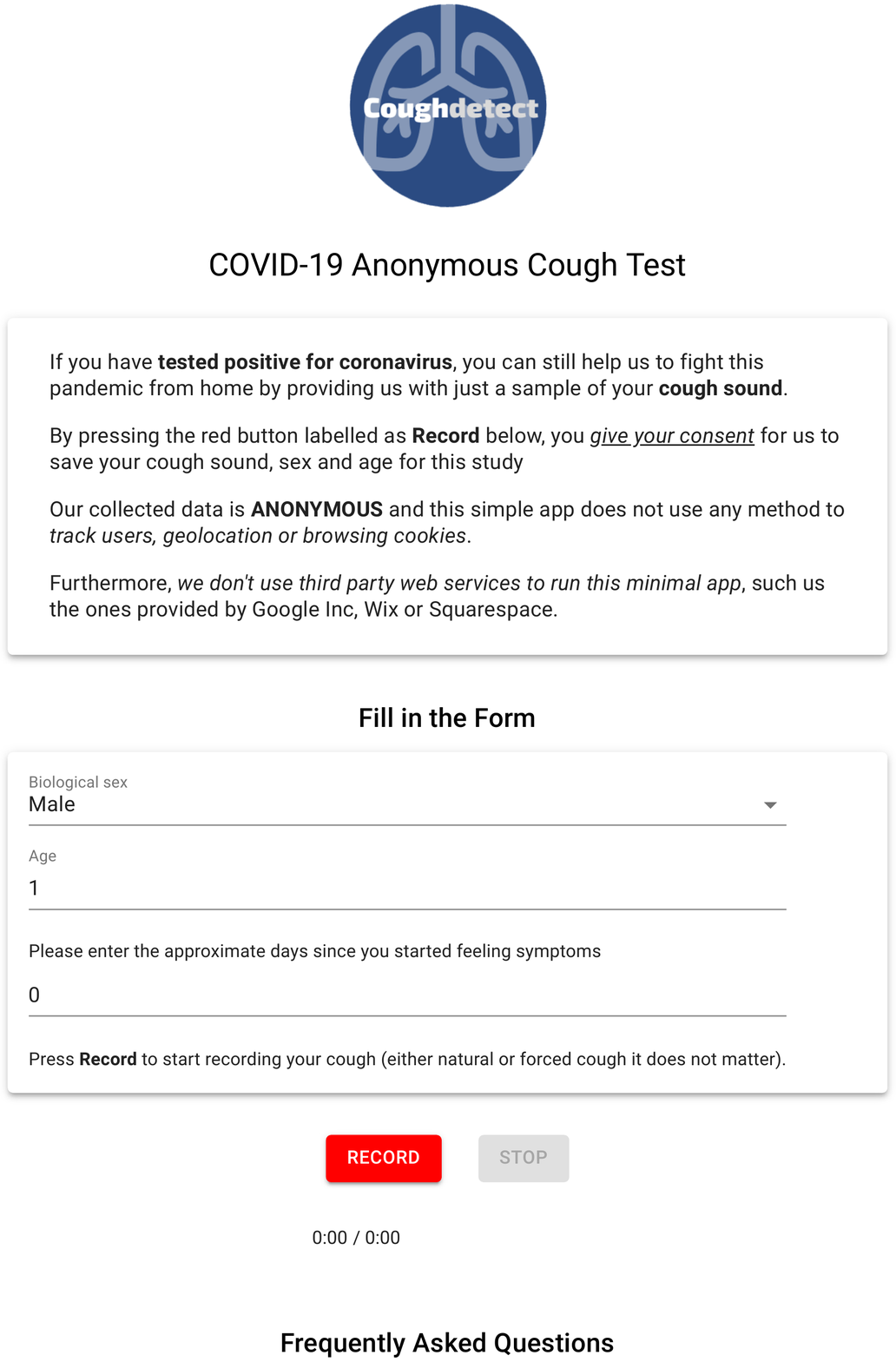}};
\node (CoughSample) [decision, text width=2cm, fill= red!20, right= of App_screenshot, node distance = 1cm, xshift=-2cm] {\large Cough Detected?};
\node (InfMech) [block, text width=4cm, fill= yellow!20, right= of CoughSample, yshift = 2cm] {\Large \textbf{DeepCough}};
\node (Mesg2user_1) [block, text width=4cm, fill= blue!20, right= of CoughSample, yshift = -2cm] {CoughDetect: \texttt{       Cough not detected, please try again.}};
\node (Mesg2user_2) [block, text width=4cm, fill= blue!20, right= of InfMech, yshift = 2cm] {CoughDetect:\texttt{  Covid-19 likely\(^*\).}};
\node (Mesg2user_3) [block, text width=4cm, fill= blue!20, right= of InfMech, yshift = -2cm] {CoughDetect:  \texttt{  Covid-19 not likely\(\dagger\).}};
\draw [myarrows,<->] (FEnd) -- (BEnd);
\draw [myarrows,<->] (BEnd) -| (PS);
\draw [myarrows,<->] (BEnd) -| (MDB);
\draw [myarrows] (App_Arch) -- (App_screenshot);
\draw [myarrows] (App_screenshot) -- (CoughSample);
\draw [myarrows] (CoughSample) |- node[xshift = 1cm]{Yes} (InfMech);
\draw [myarrows] (CoughSample) |- node[xshift = 1cm]{No} (Mesg2user_1);
\draw [myarrows] (InfMech) |- (Mesg2user_2);
\draw [myarrows] (InfMech) |- (Mesg2user_3);
\end{tikzpicture}}
\caption{A user can record his cough sample using the CoughDetect web or mobile app with complete anonymity. The user's cough sample is then analysed by DeepCough (the inference mechanism of CoughDetect) for primary screening of Covid-19. A user can receive one of the following two messages on successful analysis of his cough sample: \small \(^*\)\texttt{Your cough sound shares similarities to those of Covid-19 patients, if you are a high-risk individual, please contact health services immediately, otherwise quarantine yourself}. \(^\dagger\)\texttt{Our system does not recognise your pattern as similar to those with Covid-19 in our database, still if you feel the most likely symptoms, please contact health services.\\}}
\label{fig:CoughDetectApp}
\end{figure*}
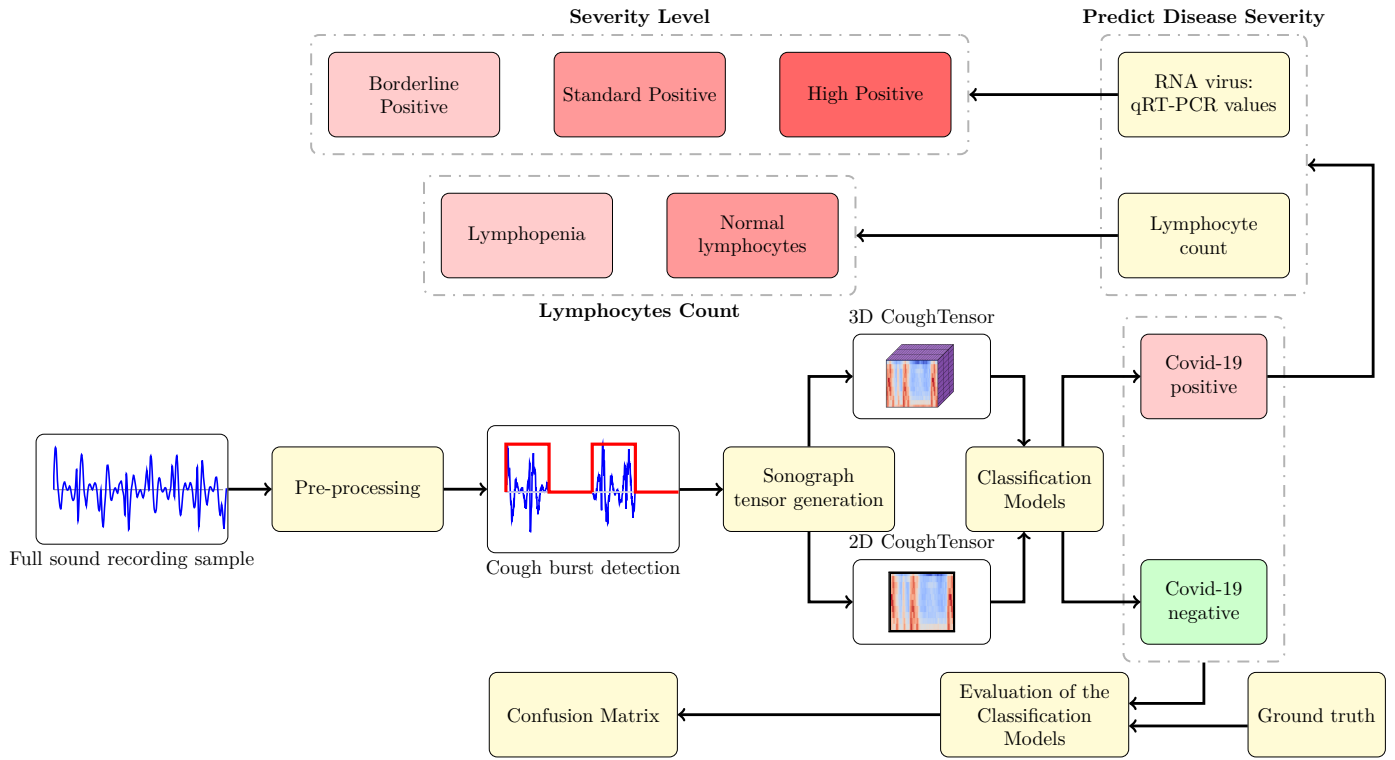
\begin{figure*}[!hbt]
\centering
\scalebox{0.75}{
\begin{tikzpicture}[node distance = 4cm, auto]
\node [block, draw,  label=below:Full sound recording sample, inner ysep=2mm,inner xsep=3mm] (cough_waves){\usebox\audiosignal};
\node [block, right of = cough_waves, fill= yellow!20] (preprocessing) {Pre-processing};
\node [block, right of = preprocessing, label=below:Cough burst detection, inner ysep=3mm,inner xsep=3mm] (seg_cough_waves){\usebox\segaudiosignal};
\node [block, right of = seg_cough_waves, fill= yellow!20] (feature_tensor) {Sonograph tensor generation};
\node [block, right of = feature_tensor, fill= yellow!20, text width=2.2cm] (classification) {Classification Models};
\node [block, above of = feature_tensor, xshift = 2cm, node distance = 2cm,  text width=2.2cm, label = 3D CoughTensor] (DC3D) {\includegraphics[scale = 0.1, trim = 10 10 10 10, clip = true]{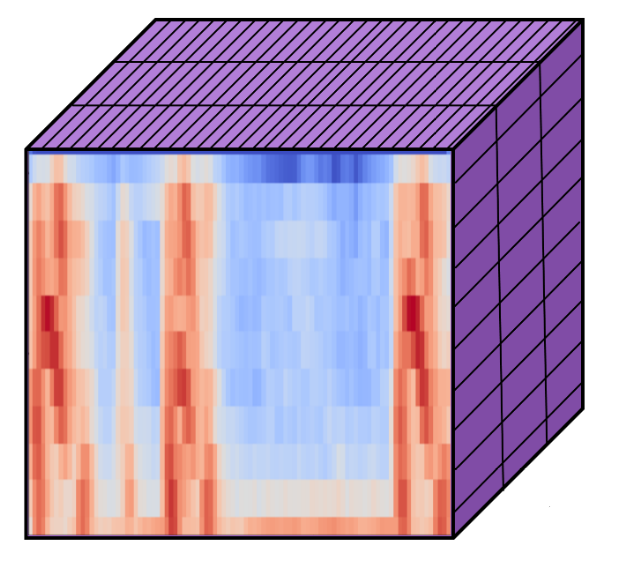}};
\node [block, below of = feature_tensor, xshift = 2cm, node distance = 2cm, text width=2.2cm, label = 2D CoughTensor] (DC2D) {\fcolorbox{black}{black}{\includegraphics[scale = 0.2, trim = 10 10 10 10, clip = true]{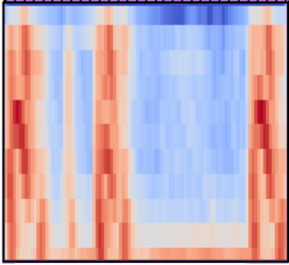}}};
\node [block, right of = classification, fill= red!20, yshift = 2cm, text width = 2cm, node distance = 3cm] (pos) {Covid-19 positive};
\node [block, right of = classification, fill= green!20, yshift = -2cm, text width = 2cm, node distance = 3cm] (neg) {Covid-19 negative};
\node [dotted_block, fit = (pos) (neg), inner ysep=3mm,inner xsep=3mm] (pos_neg) {};
\node [block, below of = classification, fill= yellow!20, yshift = -2cm, text width = 3.1cm, node distance = 2cm] (eval) {Evaluation of the Classification Models};
\node [block, right of = eval, fill= yellow!20, text width = 2.2cm, node distance = 5cm] (truth) {Ground truth};
\node [block, left of =eval, fill= yellow!20, text width = 3.1cm, node distance = 8cm] (conf_matrix) {Confusion Matrix};
\node (disease_sev_1) [block,fill= yellow!20, above  of = pos, node distance = 5cm] {RNA virus: qRT-PCR values};
\node (disease_sev_2) [block,fill= yellow!20, below of = disease_sev_1, node distance = 2.5cm] {Lymphocyte count};
\node (disease_sev) [dotted_block, fit = (disease_sev_1) (disease_sev_2),inner ysep=3mm,inner xsep=3mm, label= \textbf{Predict Disease Severity}]  {};
\node (SeverityLevel1) [block,fill= red!20, left  of = disease_sev_1, node distance = 14cm] {Borderline Positive};
\node (SeverityLevel2) [block,fill= red!40, right  of = SeverityLevel1, node distance = 4cm] {Standard Positive};
\node (SeverityLevel3) [block,fill= red!60, right  of = SeverityLevel2, node distance = 4cm] {High Positive};
\node (SL) [dotted_block, fit = (SeverityLevel1) (SeverityLevel2)(SeverityLevel3),inner ysep=3mm,inner xsep=3mm, label= \textbf{Severity Level}]  {};
\node (Lymp_normal) [block,fill= red!20, left  of = disease_sev_2, node distance = 12cm] {Lymphopenia};
\node (Lymp_abnormal) [block,fill= red!40, right  of = Lymp_normal, node distance = 4cm] {Normal lymphocytes};
\node (Lymp) [dotted_block, fit = (Lymp_normal) (Lymp_abnormal),inner ysep=3mm,inner xsep=3mm, label= below:\textbf{Lymphocytes Count}]  {};
\draw [ myarrows] (cough_waves) -- (preprocessing);
\draw [ myarrows] (preprocessing) -- (seg_cough_waves);
\draw [ myarrows] ([yshift = -0.005cm] seg_cough_waves.east) -- (feature_tensor);
\draw [myarrows] ( feature_tensor.north) |- (DC3D);
\draw [myarrows] ( feature_tensor.south) |- (DC2D);
\draw [myarrows] (DC3D.east) -| ([xshift = -.5cm]classification);
\draw [myarrows] (DC2D.east) -| ([xshift = -.5cm]classification);
\draw [myarrows] ([xshift = .5cm]classification.north) |- (pos);
\draw [ myarrows] ([xshift = .5cm]classification.south) |- (neg);
\draw [ myarrows] (pos_neg) |- ([yshift=0.2cm] eval.east);
\draw [ myarrows] ([yshift=-0.2cm]truth.west) -- ([yshift=-0.2cm] eval.east);
\draw [ myarrows] (eval) -- (conf_matrix);
\draw [myarrows] (pos) -- +(3,0) |- (disease_sev.east);
\draw [myarrows] (disease_sev_1) -- (SL);
\draw [myarrows] (disease_sev_2) -- (Lymp);
\end{tikzpicture}}
\caption{The overall flow diagram delineating the steps involved in the DeepCough, 2D and 3D, inference mechanism.}
\label{fig:COVIDflowchart}
\end{figure*}
\definecolor{GreenOlive}{rgb}{0.33, 0.42, 0.18}
\begin{figure*}[h]
    \centering
    \adjustbox{valign=t}{
    \begin{tabular}{c c c}
         \multirow{2}{*}[1.2in]{\includegraphics[clip, trim = 2cm 1cm 1.85cm .5cm, scale =0.4]{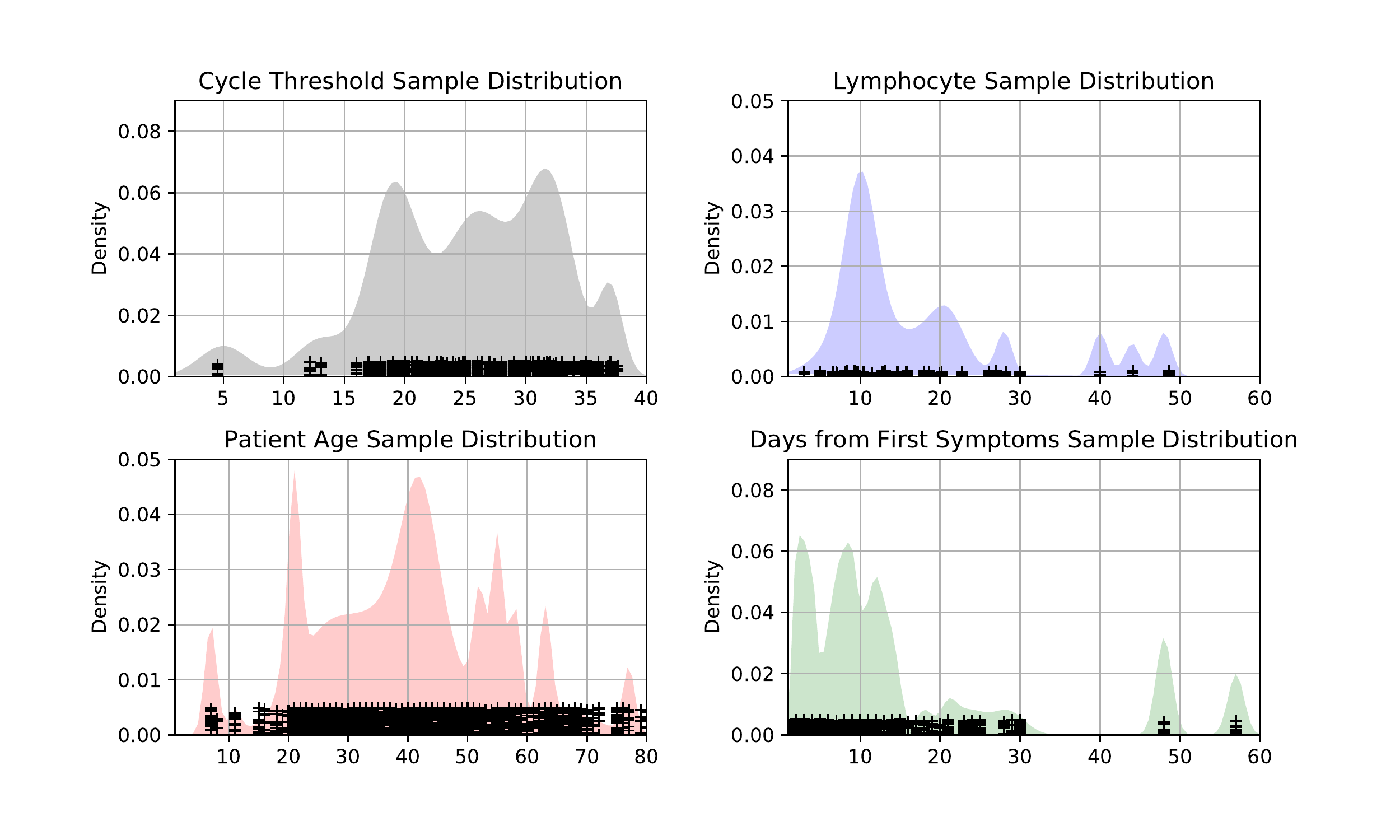}} & 
         \scalebox{0.35}{\hspace{-1cm} \vspace{-.25cm}\begin{tikzpicture}
            \pie [rotate = 180, explode = 0.1]
            {37.47/\Large \textbf{Female},
            60.20/\Large \textbf{Male}, 
            2.33/\Large \textbf{Not Specified}}
            \end{tikzpicture}} 
            &  \hspace{-.1cm}\vspace{-.25cm} \scalebox{0.35}{\begin{tikzpicture}
        \pie [color={red!40, blue!40, yellow!40},rotate = 180, explode = 0.1]
         { 37.58/ \Large \textbf{Female},
           60.97/ \Large \textbf{Male},
           1.45/ \Large \textbf{Not Specified}}\end{tikzpicture}}\\
            &  \scalebox{0.35}{\hspace{-1cm}\begin{tikzpicture}
        \pie [rotate = 180, explode = 0.1]
         { 35.14/ \Large \textbf{Borderline Positive},
           45.95/ \Large \textbf{Standard Positive},
           18.92/ \Large \textbf{High Positive}}
        \end{tikzpicture}}&\hspace{-.1cm} \includegraphics[clip, trim = 0cm 0cm 0cm 0cm, scale =0.3]{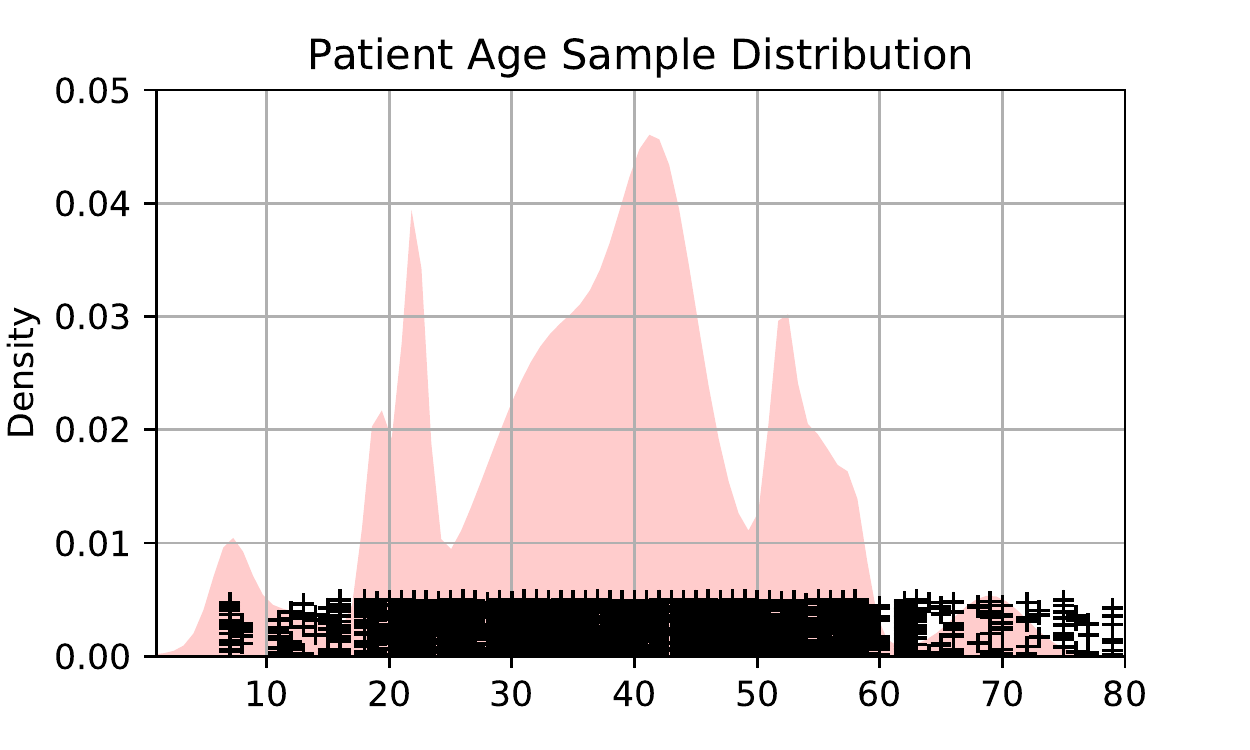}
        \\
        a) Density Distributions for positive samples. & b) Positive samples & c) Negative samples\\
    \end{tabular}}
    \caption{a) Density distributions of cycle threshold (Ct), lymphocyte count, age, and days from first symptoms from the samples of Covid-19 positive patients. b) Percentage ratios of sex (Male, Female, and Not Specified)  and level of positivity (Borderline Positive, Standard Positive, and High Positive) of samples from positive Covid-19 patients displayed in pie charts. c) Percentage ratios of sex (Male, Female, and Not Specified) of samples from negative Covid-19 displayed in a pie chart along with density distribution of age for Covid-19 negative patients.}
    \label{fig:distribution}
\end{figure*}


\subsection{Research Ethics and Protocol}
The collection of clinically validated cough data was carried out in collaboration with \emph{Hospital Costa del Sol Health Agency} in M\'alaga, Spain and the \emph{National Laboratory for Research in Food Safety (LANIIA)} laboratory in Nayarit, Mexico. The collection of the data started at the peak of the pandemic in Spain and Mexico on the 4th of April 2020 and lasted until the 21st of September 2020. The clinical protocols and research ethics are approved by the respective local institutional ethics committees (Code: BIOETIC\_HUM\_2020\_02, Mexico; Code: APP\_Covid\-19\_03042020, Spain). The Nayarit Unit and M\'alaga hospital are both accredited centres for the molecular diagnosis of Covid-19 and are also ISO 9001 certified.

The cough samples are collected from patients coming to the named institutions for a qRT-PCR test for detection of SARS-Cov-2 (Covid-19) by registered nurses trained to use the CoughDetect app. At all stages of the cough sample collection, the guidelines to interact with potential Covid-19 patients recommended by the WHO are strictly followed. For instance, the nurse wears personal protective equipment at all times, and a protocol for the smartphone disinfection, each time a cough is recorded, is observed.

The user interface and control functions of the Web App have been developed with in-house code to uphold the anonymity of the users and minimise the possibility of information leakages to external entities. This is in conformity with both the EU General Data Protection Regulation (GDPR) and the UK Data Protection Act 2018. In addition, our research and application also meet the ethical standards of the Declaration of Helsinki. A written informed consent was collected from each participant prior to  acquiring their data sample. Clinical data was collected for this study by healthcare professionals. Table \ref{tab:desc_stat_negpos}, Fig. \ref{fig:distribution} summarises the demographic ratios and factors such as the number of days after first symptoms reported. In total, we collected $n=8,380$ coughs, of which 2,339 coughs are from patients with a positive qRT-PCR test and 6,041 coughs are from patients with a negative qRT-PCR test. Of those patients who resulted negative in the qRT-PCR test, 47.46\% had no symptoms, and 52.54\% had symptoms, at the time of taking the samples. Of those patients who resulted positive in the qRT-PCR test, 20.00\% had no symptoms, and 80.00\% had symptoms, at the time of taking the samples.

\begin{table}[!b]
 \caption{Demographic statistics of the data (Covid-19 positive and negative patients)}
 \label{tab:desc_stat_negpos}
 \centering
 \scalebox{0.85}{
 \begin{tabular}{|l|l|l|l|l|l|l|}
 \hline
&
\multicolumn{3}{c|}{Positive} &
\multicolumn{3}{c|}{Negative}\\
\hline
 \textbf{Measure}   &  \textbf{Age}   & \textbf{Days$^{*}$}  & \textbf{PCR CT} &  \textbf{Age}   & \textbf{Days$^{*}$}  & \textbf{PCR CT}     \\\hline
 \textbf{Mean}      & 39.44  & 7.74 & 29.21  & 38.74  & 7.90 & 40.23  \\\hline 
 \textbf{Median}    & 38     & 7     &  31   & 38  & 6 & 41 \\\hline
 \textbf{Std. Dev.} & 14.24  & 6.39  &  7.13 & 13.59  & 6.67 & 6.16\\\hline
 \textbf{Max}       & 79     & 60    &  37   & 79  & 50 & 43\\\hline
 \textbf{Min}       & 7      & 1     &  18   & 7  & 0 & 38\\\hline 
  \end{tabular}}
\footnotesize{$*$number of days since the onset of symptoms}
\end{table}

\subsection{Cough sound pre-processing and detection}
\label{section:preprocessing}
Cough samples (\texttt{.wav}) are were acquired at \texttt{44.10 kHz}, Pulse-code modulation (PCM) format, monochannel. The raw sound data is low pass filtered with a cutoff frequency of \texttt{1 kHz}. A Chebyshev type-2 second order filter with a transition frequency of \texttt{10Hz} is applied to retain the high pitch sound of cough while attenuating background sounds simultaneously. Before cough detection, the filtered sound signal is decimated. For an initial bout of sounds in the recording, such as initial involuntary voice before coughing, envelope analysis detects the first peak amplitude and the following signal is subsequently trimmed.

The cough detection algorithm with the filtered audio signals is based on empirical mode decomposition (EMD) \cite{emd98,emd03}. EMD is a fully data-driven signal processing technique that does not employ base functions. EMD splits a sequence into a set of smaller sequences, referred to as intrinsic mode functions (IMFs), or simply modes, whereby each mode contains the energy associated at a certain scale. EMD has become popular in many applications, e.g. wearable sensors \cite{bsn20}, perhaps because the decomposition occurs in the same space as the original sequence.

\begin{figure*}[t]
\centering
\scalebox{0.7}{
\begin{tikzpicture}[node distance = 4.5cm, cross/.style={path picture={ 
  \draw[black]
(path picture bounding box.south) -- (path picture bounding box.north) (path picture bounding box.east) -- (path picture bounding box.west);
}}]
\node (downsampling) [block, fill= yellow!20]  {Downsampling by a factor of 10};
\node (Step1_DS_fig) [below of =downsampling, node distance = 5cm]  {\includegraphics[clip, trim=1cm .5cm 0.5cm .1cm, scale=0.415] {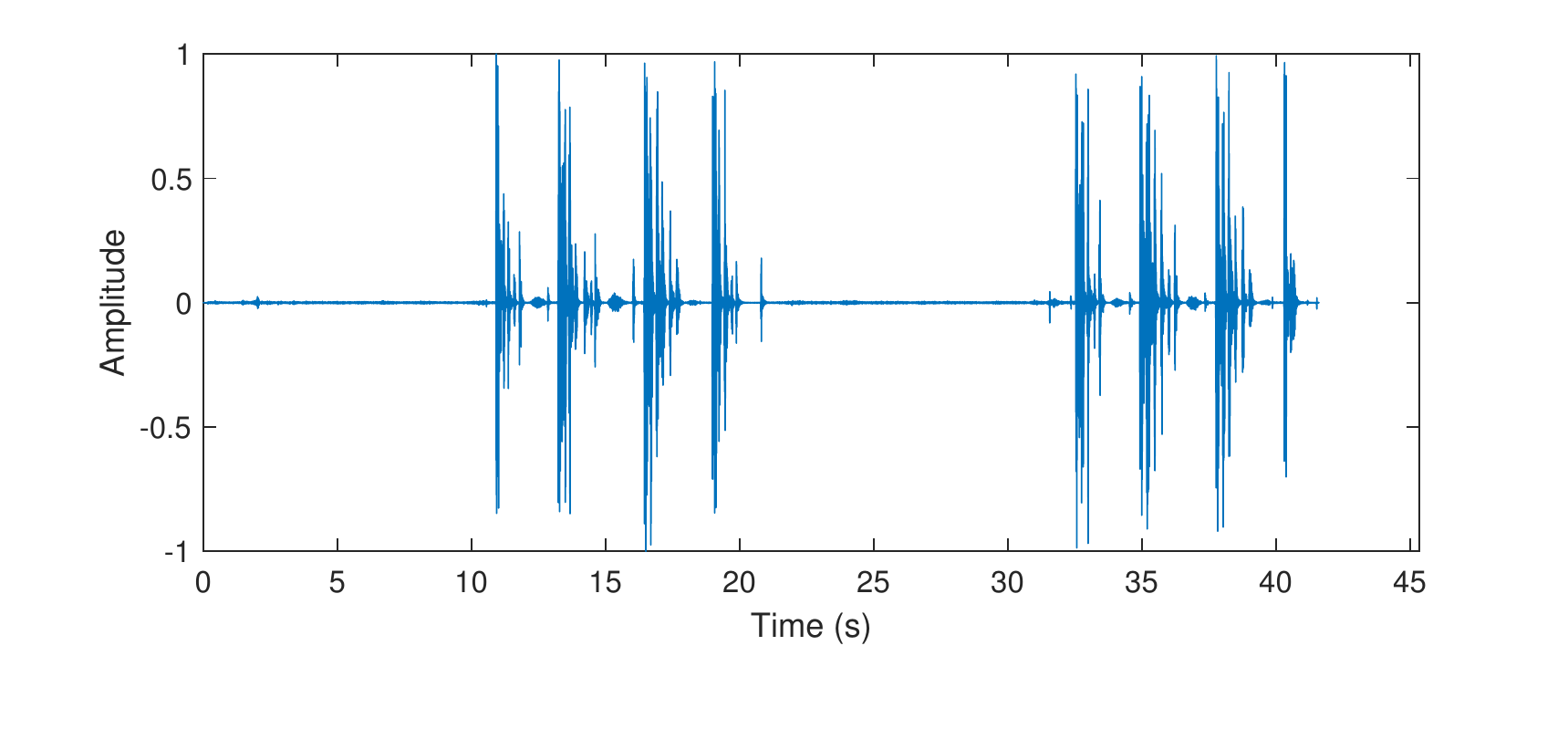}};
\node (Step1_DS_block) [dotted_block, fit = (Step1_DS_fig), label=below: \textbf{Downsampled Audio Signal}]  {};
\node (EMD) [block, right of = downsampling, fill= yellow!20, text width = 3.5cm]  {Empirical Mode Decomposition (EMD)};
\node (Step2_EMD_fig) [above of =EMD, node distance = 5cm, xshift = -4.5cm]  {\includegraphics[clip, trim=1cm 0.7cm .7cm .2cm, scale=0.5] {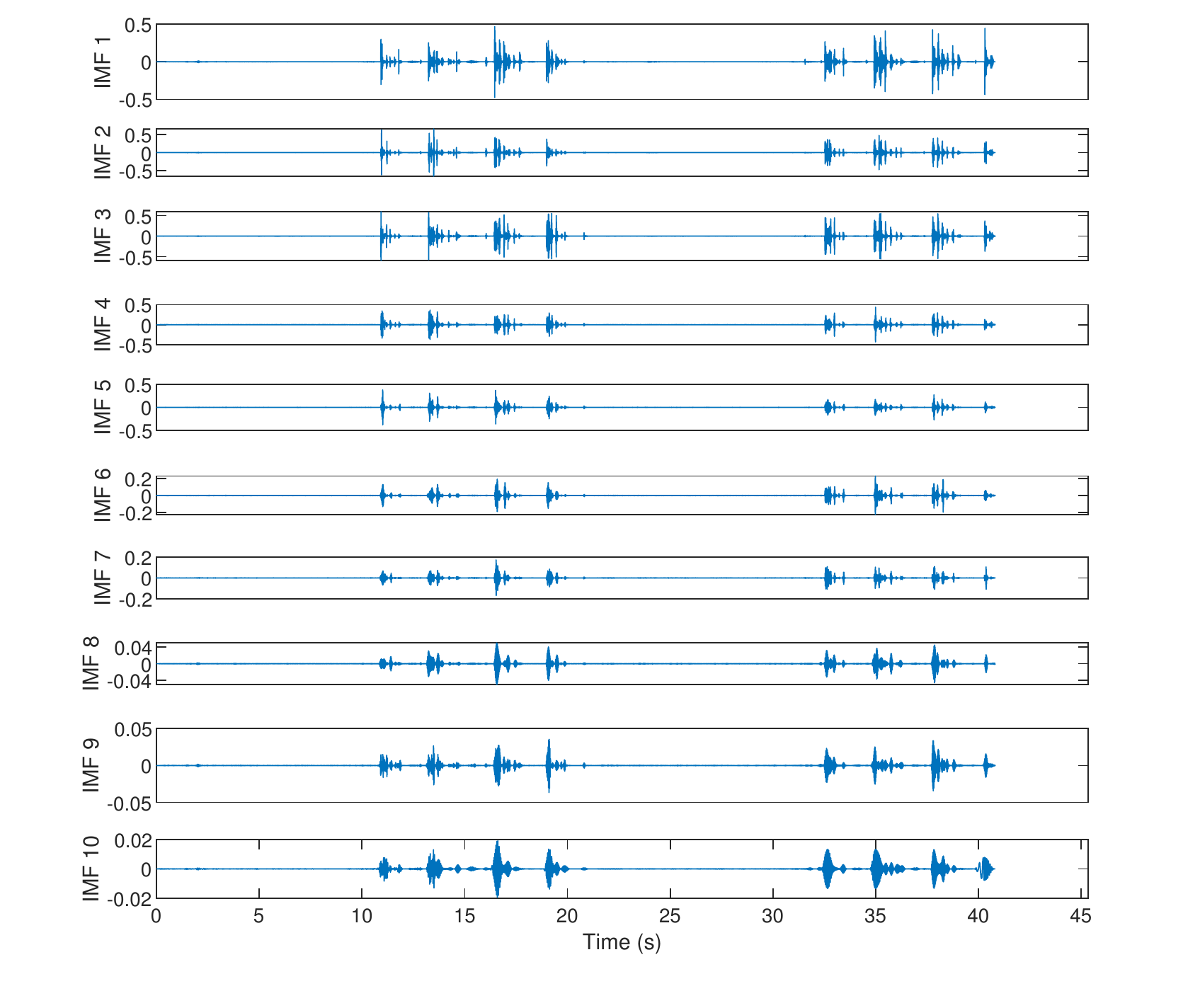}};
\node (Step2_EMD_block) [dotted_block, fit = (Step2_EMD_fig), label=above: \textbf{EMD}]  {};
\node (IMF) [block, right of = EMD, fill= yellow!20, text width = 3cm]  {Intrinsic Mode Function (IMFs) Selection};
\node (Step3_IMF_fig) [below of =IMF, node distance = 5cm]  {\includegraphics[clip,  trim=1.1cm .2cm 1cm .5cm, scale=0.5] {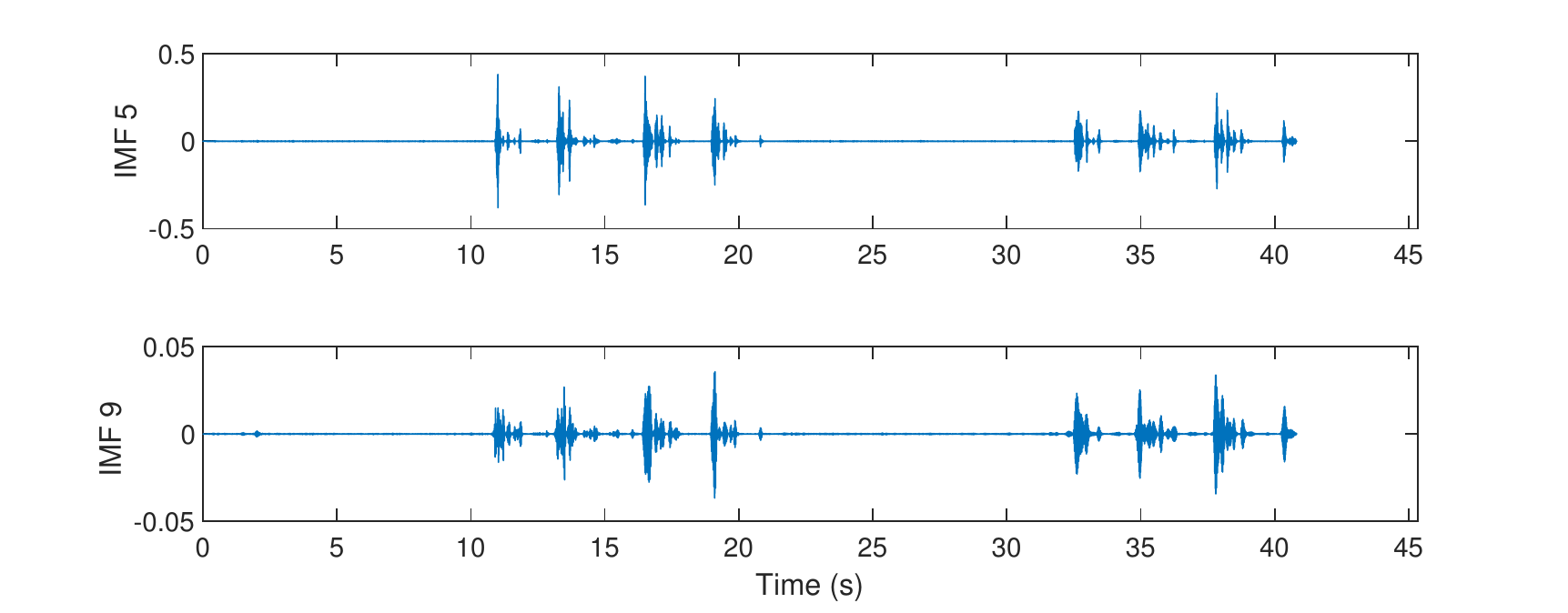}};
\node (Step3_IMF_block) [dotted_block, fit = (Step3_IMF_fig), label=below: \textbf{IMF Selection}]  {};
\node (HT) [block, right of = IMF, fill= yellow!20, text width = 3cm]  {Hilbert Transform (HT)};
\node (Step4_HT_fig) [above of =HT, node distance = 5cm, xshift = -4.5cm]  {\includegraphics[ clip,  trim=1.1cm .2cm .5cm .1cm, scale=0.5] {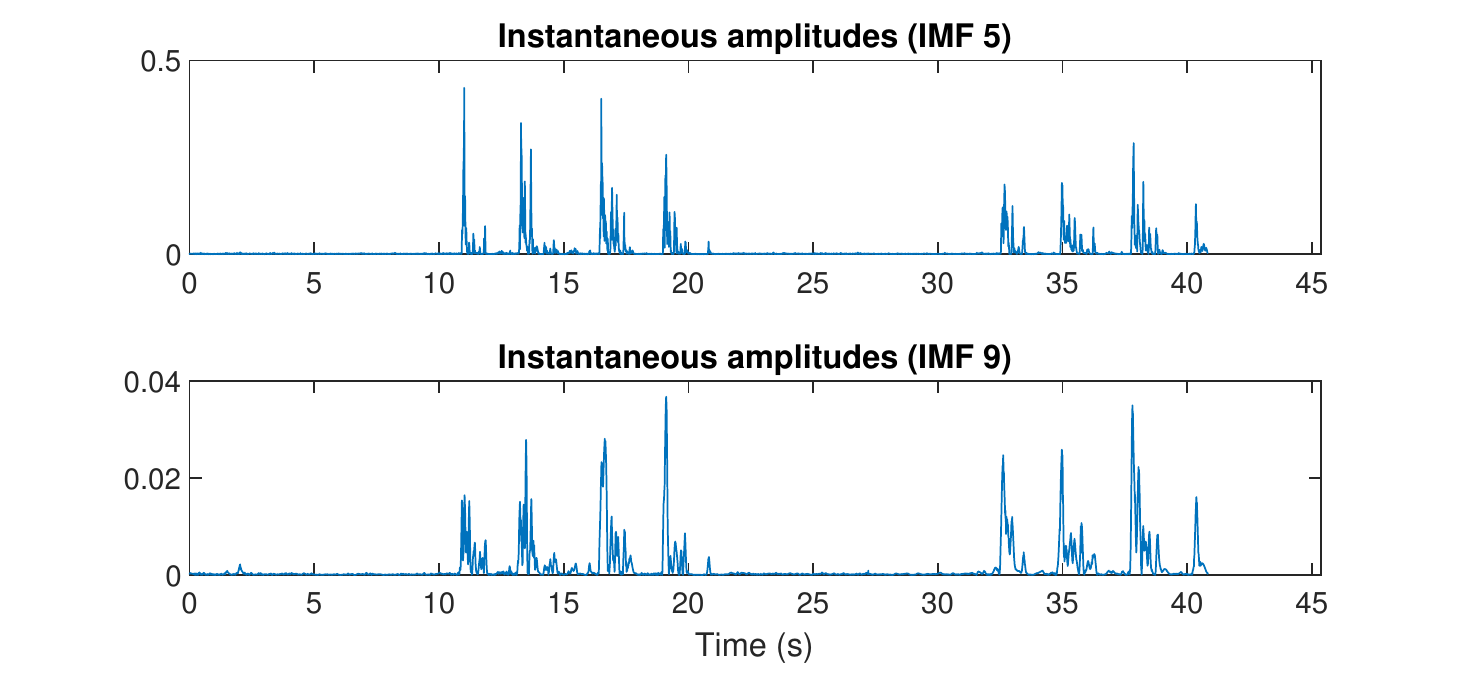}};\node (Step4_HT_block) [dotted_block, fit = (Step4_HT_fig), label=above: \textbf{Hilbert Amplitudes}]  {};
\node (circle_add)[draw,thick, circle,cross,minimum width=.3cm, right of= HT, node distance = 3.5cm]{};
\node (Step5_MF_fig) [below of =circle_add, node distance = 5cm, xshift = 0.2cm]  {\includegraphics[clip, trim=1.1cm .2cm .5cm .1cm, scale=0.41] {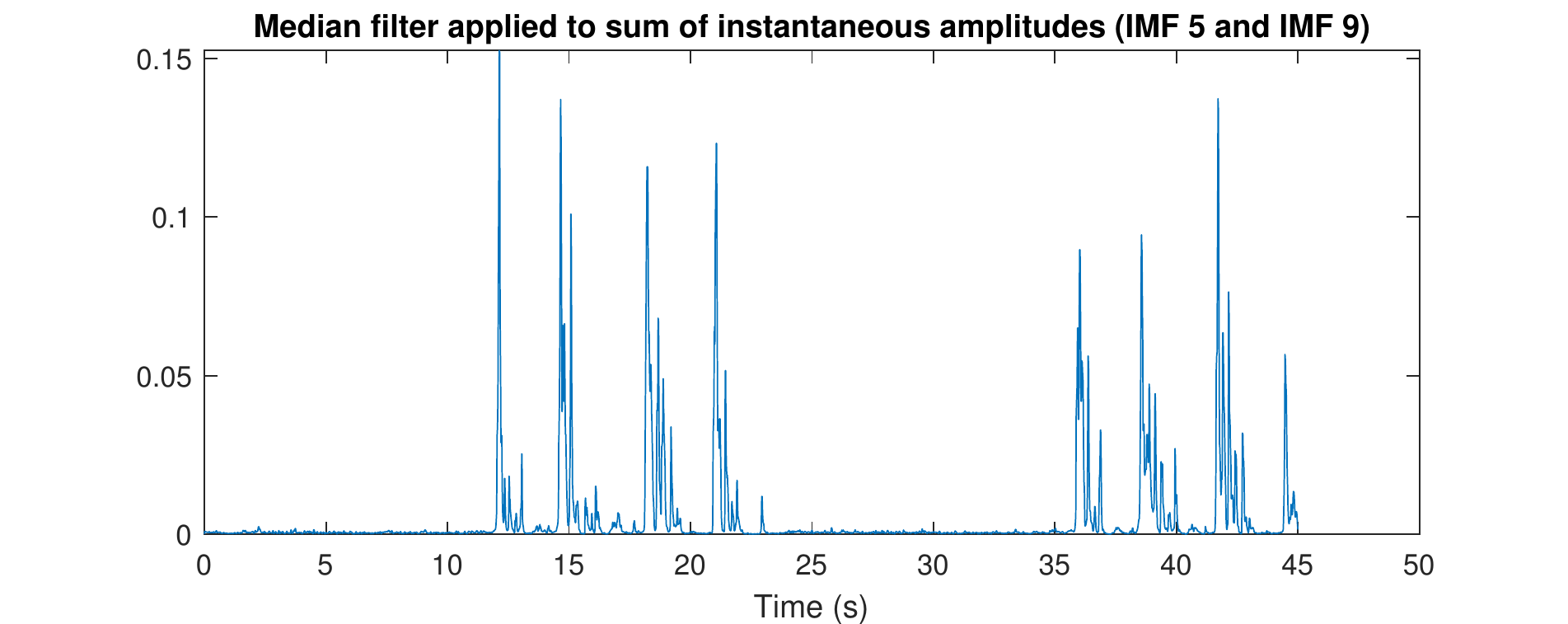}};
\node (Step5_MF_block) [dotted_block, fit = (Step5_MF_fig), label=below: \textbf{Median Filter}]  {};
\node (Step6_Cough_fig) [above of =circle_add, node distance = 5cm, xshift = 1cm]  {\includegraphics[clip, trim=1.1cm .2cm .5cm .1cm, scale=0.4] {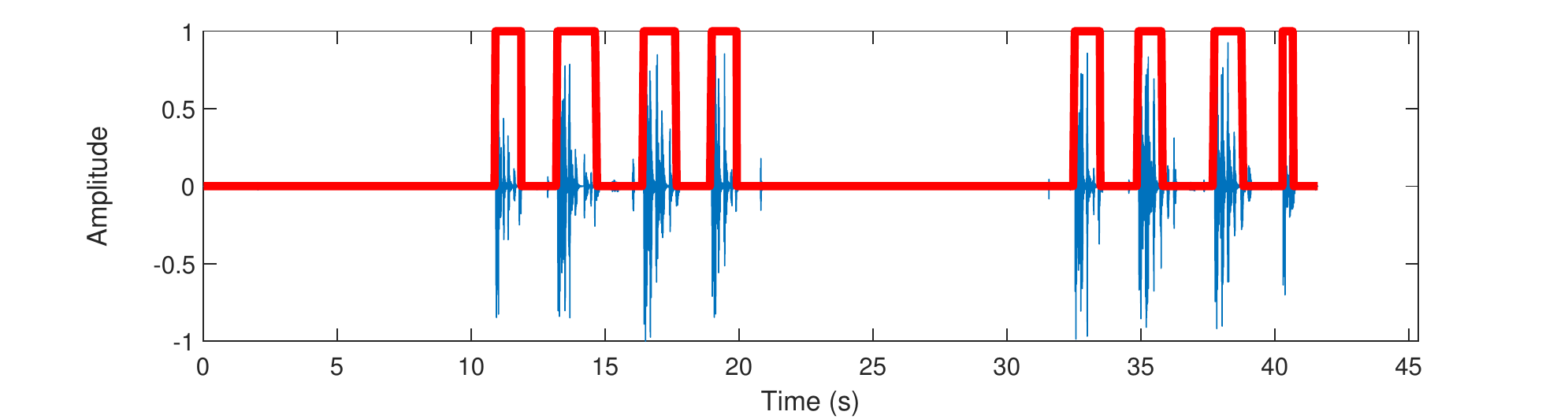}};
\node (Step6_Cough_block) [dotted_block, fit = (Step6_Cough_fig), label=above: \textbf{Detect cough burst}]  {};
\draw [myarrows](-3,0) -- node[font=\small, text width = 1cm, yshift = 0.5cm]{Input Signal}(downsampling.west);
\draw [myarrows] (downsampling) -- (EMD);
\draw [myarrows] (EMD) -- (IMF);
\draw [ myarrows] ([yshift = 0.3cm]IMF.east) --node[font=\small, text width = 1.2cm, yshift = 0.4cm, xshift = 0.05cm]{IMF \# 1 (5th)} ([yshift = 0.3cm]HT.west);
\draw [ myarrows] ([yshift = -0.3cm]IMF.east) --node[font=\small, text width = 1.2cm, yshift = -0.4cm, xshift = 0.05cm]{IMF \# 2 (9th)} ([yshift = -0.3cm]HT.west);
\draw[myarrows] ([yshift = 0.5cm]HT.east) -| node[font=\small, text width = 2cm, yshift = 0.4cm, xshift = -.8cm]{Instantaneous Amplitude 1}(circle_add.north) ;
\draw[myarrows] ([yshift = -0.5cm]HT.east) -| node[font=\small, text width = 2cm, yshift = -0.4cm, xshift = -.8cm]{Instantaneous Amplitude 2}(circle_add.south) ;
\draw[myarrows] (circle_add.east) -- node[font=\small, text width = 2cm, yshift = 0.2cm, xshift = .1cm]{Thresholding} ++(2,0) |- node[font=\small, text width = 1cm, yshift = 0.2cm, xshift = .5cm]{Cough}++(1,1);
\draw[myarrows] (circle_add.east)-- ++(2,0) |- node[font=\small, text width = 1.5cm, yshift = -0.4cm, xshift = .8cm]{Non-Cough}++(1,-1);
\draw[dashedline] (downsampling) -- (Step1_DS_block);
\draw[dashedline] (EMD) |- (Step2_EMD_block);
\draw[dashedline] (IMF) -- (Step3_IMF_block);
\draw[dashedline] (HT) |- (Step4_HT_block);
\draw[dashedline] ([xshift=0.2cm]circle_add.east) -- ([xshift=0.2cm]Step5_MF_block.north);
\draw [dashedline] ([xshift =2cm]Step6_Cough_block.south) -- ++(0,-2.5);
\end{tikzpicture}}
\caption{A pictorial illustration of the steps involved in the detection algorithm.}
\label{fig:segmentation}
\end{figure*}
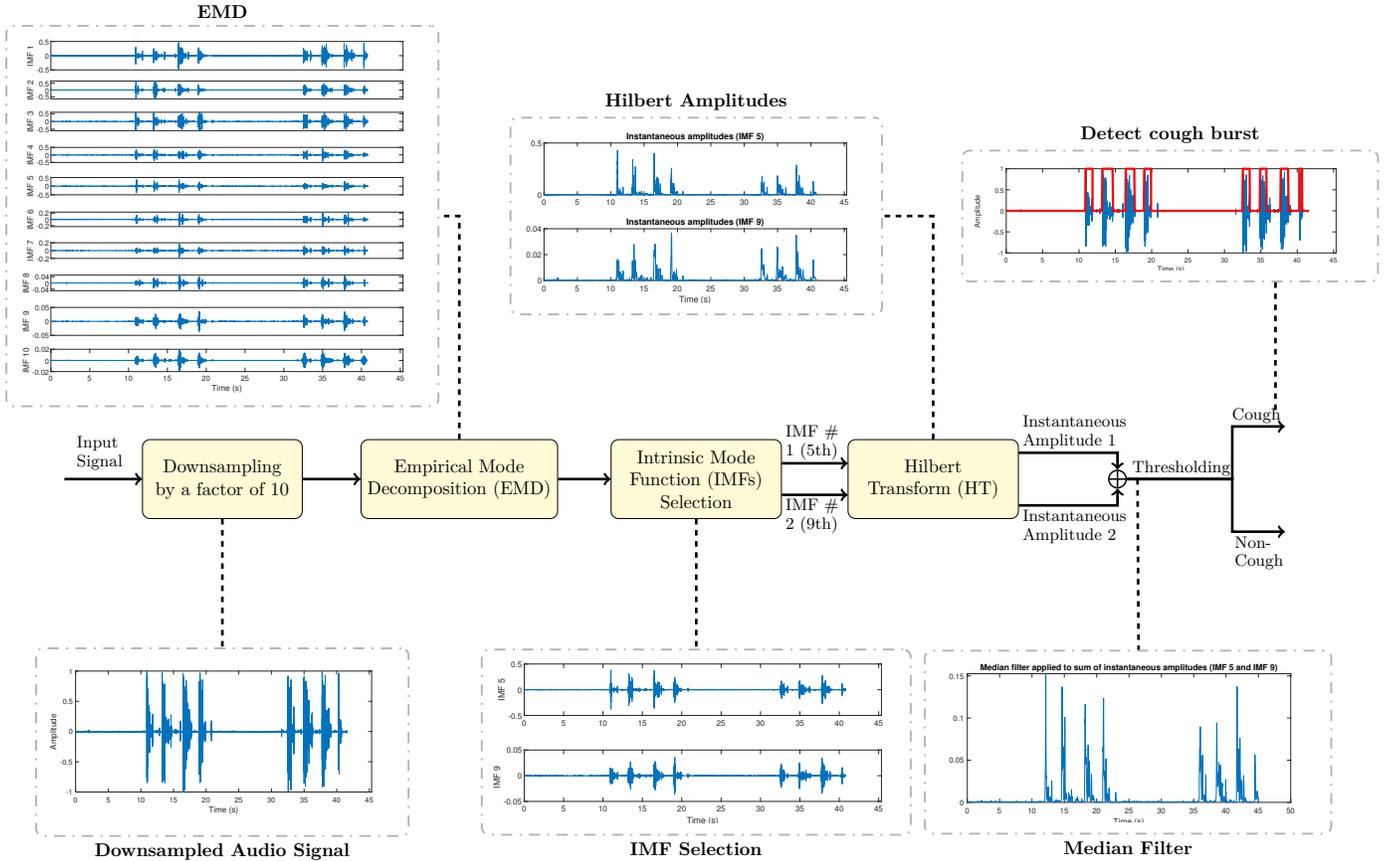

EMD is applied to find the modes that better reflect the coughing periods. These periods are empirically selected to essentially detect cough burst in the filtered sound recordings. Individual or a set of IMFs can be objectively used for signal filtering, peak detection and signal reconstruction. For cough detection, depending on the noise level of the signal, certain IMFs contain rich information related to the peaks associated with coughs. Based on testing a number of signals with various noise levels, the $5^{th}$ and the $9^{th}$ modes are found as the prime IMFs essential for detection.

The instantaneous amplitudes (IAs) of the selected modes ($5^{th}$ and $9^{th}$) are calculated by the Hilbert Transform \cite{emd98}. The IAs of the selected modes are averaged, 
low pass filtered using a median filter with a window size of 500 signal samples, and normalised. Thresholding is performed using local signal peak detection: A signal sample is a local mode or peak if it has the local maximal value being preceded (to the left) by a value difference of $\Delta \ge 0.006$. Thresholding the processed IAs partitions the original signal into cough and non-cough burst event. A summary of the EMD based algorithm for cough detection is depicted in Fig.~\ref{fig:segmentation}.

The detection algorithm produces a sequence of binary values: ones for cough and zeros for non-cough segments. A post-processing step joins consecutive cough bursts (segments) which are part of a single or main cough. To do this, an additional threshold is specified to decide whether to join neighbouring cough bursts with a distance shorter than 1500 decimated signal samples (0.34 seconds). Once an entire cough sound is detected, the rest of the signal is discarded. In addition, segments of short duration (length less than 400 signal samples) were discarded as they were often found to be more representative of short spikes in the signals due to ambient noise rather than part of a cough sound. The final output is a vector of indices that indicates where a cough in a raw sound stream is found.
\subsection{A CoughTensor of Sonographs}
\label{section:tensor}
Following detection, the information contained in the audio signals is transformed into a tensor form. We focused on representations that capture the main acoustic properties of the coughs. We used three types of \emph{sonographs}: 1) Mel-frequency Cepstral Coefficients (MFCCs), 2) Mel-scaled spectrogram (MelSpec), and 3) Linear Predictive Coding Spectrum (LPCS) coefficients. These sound representations have specific properties for classification in intelligent audio analysis. We describe them here and discuss what they can inform us about coughs sounds.

\subsubsection{Mel-frequency Cepstral Coefficients}
\emph{MFCCs} take into account human auditory perception, where low frequencies are better understood than high frequencies. The frequency bands are logarithmically located according to the Mel scale, which simulates the human auditory response more appropriately than the linearly spaced bands while at the same time disregards all other information. This descriptor is robust to variations in speech across subjects as well as the variations in recording conditions. MFCCs have been widely used in frequency domain speech recognition \cite{shi2018hidden, do2009recognition, al2017enhanced}. The computation of MFCCs involves the following main steps: (i) blocking of pre-processed cough sounds into overlapping windows to avoid loss of information at the ends of windows, (ii) applying hamming window on each frame to taper ends of a frame to zero so that spectral leakage can be avoided during the implementation of Fourier Transformation (FT), and (iii) computation of the power spectrum by applying FT. Next, (iv) the computed spectrum is passed through Mel-spaced band pass filters, where each filter provides the sum of energy for each frame. Finally, (v) the application of discrete cosine transformation yields MFCCs.

\subsubsection{Mel-scaled Spectrogram} 
\begin{figure}[!b]
    \centering
\includegraphics[width=0.48\textwidth]{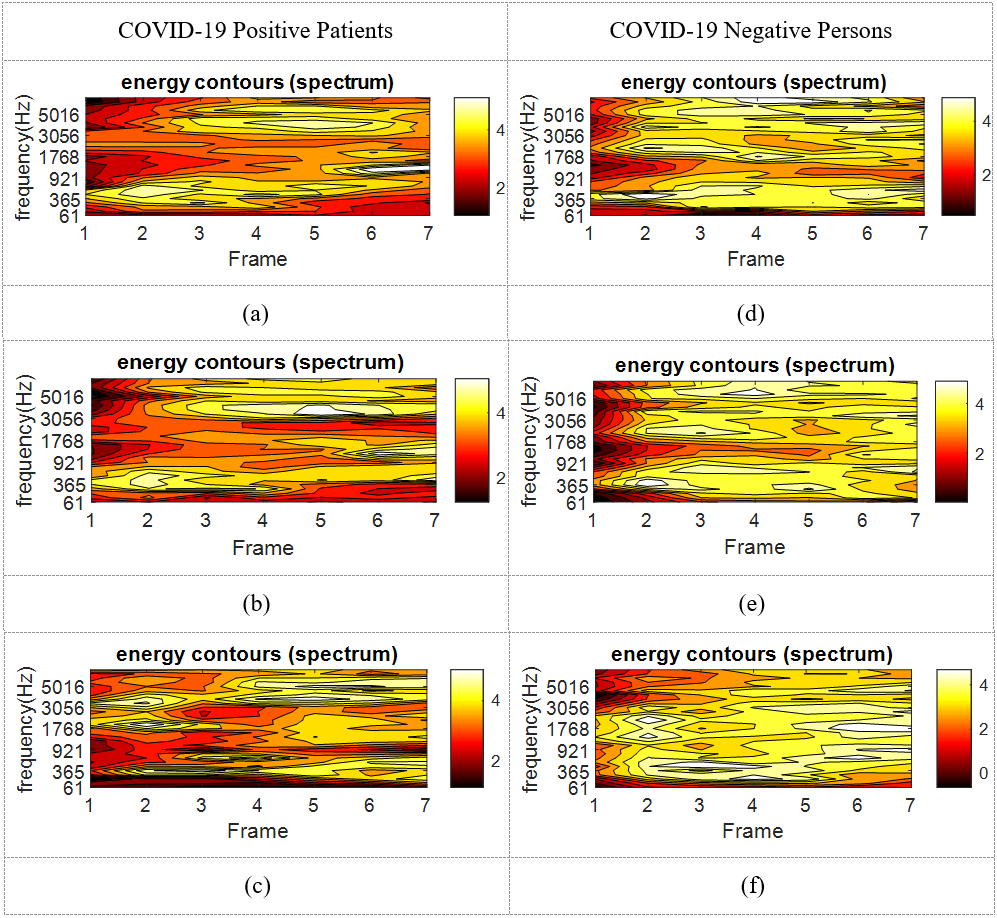}
    \caption{Energy contours of the Auditory Processed Spectrum (APS) representation which is related to MFCCs \cite{ali2016automatic} (a-c) Covid-19 positive patients (d-f) Covid-19 negative persons. Darker colors represent lower energy in the spectrum, while lighter color means higher energy.}
    \label{fig:energyDis}
\end{figure}

The \emph{MelSpec} is a sonograph where frequencies are converted to the Mel scale in order to visually assess the energy distribution in the signal. The distribution of the energy in the Mel-based spectrum is relevant for the detection of Covid-19 positive samples. Fig. \ref{fig:energyDis} provide examples of the energy spectrum for positive and negative samples. It can be observed that when a Covid-19 patient starts coughing, the energy is in the low-frequency region. However, over time the energy shifts to the high-frequency region. The lower frequencies at the start may be due to pain, and later perhaps the extra efforts required for coughing make the signal more irregular and complex over time. A similar trend is also observed in the voice of people who are suffering from pain due to vocal folds disorders. Extra efforts in speaking render the signal complex which result in an irregular spectrum (continuous voice breaks and disperses energy) compared to a healthy person \cite{ali2018intelligent, ali2016automatic}. In contrast, for a Covid-19 negative person, the energy is uniformly distributed among all frequencies. Therefore, the stark differences in MelSpecs from Covid-19 positive and negative individuals can be leveraged for successful identification of Covid-19 infection.

\subsubsection{Linear Predictive Coding Spectrum (LPCS) coefficients} LPCS models the emission source of an acoustic signal. LPCS is based on the source-filter model of phonatory signals. It is frequently used for the processing of speech and infant cry. Linear predictive coding analysis estimates the values of a signal as a linear function of previous samples. LPCS is a simplified vocal tract model that reflects the speech production system using a source-filter model. LPCS derives a compact representation of the spectral magnitude of brief duration signals (e.g. coughs). Its parametric analysis allows more accurate spectral resolution than the non-parametric FT when the signal is stationary for only a short time \cite{o1988linear}. This sound representation has been used for assessing the vocality of cough sounds \cite{van2002assessing} and detecting coughs from other human sounds \cite{swarnkar2013neural}.

For each audio frame, we extracted MFCCs with 33 coefficients, MelSpec with 33 bands, and LPCS with 33 line spectral pair frequencies from 33 coefficients. We obtained three matrices of 33 columns by the number of frames of the audio sample. The three sonographs are stacked to form a three-dimensional tensor. Since cough samples have different durations, they have different number of frames. For all samples, the tensor is padded with zeros to complete 100 x 33 x 3 matrices (see Fig. \ref{fig:COVIDflowchart}) to obtain matrices of the same shape before passing them to the training stage. 
We set a sampling rate of 22050 and a hop length of 512. The 100 frames are equivalent to around 2.3 seconds. We chose this tensor length as the minimum duration of a cough event after pre-processing and detection (section \ref{section:preprocessing}) falls in this range. Additionally, using this length we ensure that no spurious noise is included in the audio input.

\subsection{Classification of the CoughTensor via  Convolutional Neural Networks `DeepCough'}
\label{section:classifier}

The CoughTensor generated in Section \ref{section:tensor} are input to a stack of convolution blocks. Fig. \ref{fig:architectureCNN} illustrates the architecture of DeepCough along with the dimensions of each layer. The sonograph tensor is fed to the convolution blocks in a manner analogous to how RGB images are processed. 
The first dimension corresponds to the horizontal axis of the sonograph (time frames), the second dimension is the vertical axis (frequencies, bands, coefficients), and the third dimension is the type of sonograph. 
For comparison purposes we defined two types of DeepCough: 
\begin{enumerate}
\item \textbf{DeepCough2D}: The CoughTensor where 2D MelSpec is included in the tensor only, making a tensor spanning two dimensions (frequency and time) i.e. $100\times33\times1$. 
\item \textbf{DeepCough3D}: The CoughTensor stacks all sonographs described in section \ref{section:tensor}, with additional third dimension added for each sonograph hence rendering a tensor size of $100\times33\times3$.
\end{enumerate}

Each convolutional block is composed of the following layers:
\begin{itemize}
    \item Convolutional layers with rectifier linear units (ReLU): Convolution window is set to $2\times 2$ (height/width) and initial padding is set to the length of the input tensor. The input dimensions are row, column and channels.
    \item Max pooling layer: The pooling window is also set to $2\times 2$ for height and width.
    \item Dropout layer: A drop out level of 20\% probability in each block to deter the model from over-fitting.
\end{itemize}

This basic block is stacked four times, permitting a balance between architectural depth and complexity. The stack is followed by subsequent layers to transform the intermediate layer outputs for the final layer:

\begin{itemize}
    \item A global average pooling layer (GA): It averages all spatial dimensions of the input tensor until the spatial dimension is one.
    \item Dense layer (D): A dense layer yielding an output equivalent to the number of classes (one per class).
    \item Softmax layer: A softmax type action function that performs classification over the inputs.
\end{itemize}
Adaptive Moment Activation (ADAM) is the optimiser used to train the network with a categorical cross-entropy loss function. The evaluation metric during training for ADAM is the sum between resultant area under the curve (AUC) and balanced accuracy. The entire model is implemented in Keras \cite{chollet2018keras} with Tensorflow backend.

The remarkable classification prowess of DeepCough arises from representation learning via convolutional neural networks of the sonograph representations. It is not only an intuitive approach for the analysis of pattern singularities in cough sounds but also has the capacity to integrate information from different sonographs, therefore jointly performing pattern analysis in the information that comes to represent emission (MFCCs, MelSpec) and perception characteristics (LPCS) of sounds (section \ref{section:tensor}).

\begin{figure*}
\subfloat[The architecture of CNN.]{
\input{ConvolutionalNeuralNetwork}
} \hfill 
\subfloat[Summary of the CNN's architecture.]{
\label{fig:summaryCNN}
\scalebox{0.6}{
\adjustbox{valign=b}{
\begin{tabular}{|l| l|l|c|}
\hline
\textbf{Block Number}                    &\textbf{ Layer Name}    &   \textbf{Output Shape }           & \textbf{Learning Params.}    \\\hline
\multirow{3}{*}{\textbf{Block 1 (B1)}}   & Convolution + ReLU        &   (32, 99, 16)    & 208       \\
                                & MaxPooling   &   (16, 49, 16)  & n/a\\
                                & Dropout (0.2)      &   (16, 49, 16)     &  n/a\\\hline
\multirow{3}{*}{\textbf{Block 2 (B2)}}   & Convolution + ReLU         &   (15, 48, 32)      &   2080 \\
                                & MaxPooling   & (7, 24, 32)         &n/a\\
                                & Dropout (0.2)     &(7, 24, 32)        &0\\\hline
\multirow{3}{*}{\textbf{Block 3 (B3)}}   & Convolution + ReLU       & (6, 23, 64)         &8256      \\
                                & MaxPooling   & (3, 11, 64)         &n/a\\
                                & Dropout (0.2)      & (3, 11, 64)        &0\\\hline
\multirow{3}{*}{\textbf{Block 4 (B4)}}   & Convolution + ReLU        &(2, 10, 128)        &    32896    \\
                                & MaxPooling   & (1, 5, 128)         &n/a\\\
                                & Dropout (0.2)      & (1, 5, 128)         &n/a\\\hline
\textbf{GA}           & Global Averaging & (128) & n/a\\\hline
\textbf{D  }                     & Dense         &     (2 )&        258 \\\hline
\end{tabular}
}}}
\caption{(a) An illustration of the architecture of the Convolutional Neural Network (CNN) with (b) dimensions of convolutional blocks (B1-B4), max-pooling layers, a global averaging (GA), and a dense (D) layer.}
\label{fig:architectureCNN}
\end{figure*}


\subsection{Development and Deployment of DeepCough as an Anonymous Web-based primary screening for Covid-19}
\label{section:weapp}
The methods described in this paper are deployed in a Web App proof of concept (POC) available at \url{https://coughdetect.com}. The main objectives of the interface are as follows:

\begin{itemize}
    \item Enable a sleek and multi-platform Web App that can be accessed from any device with connectivity to the Internet without installation i.e. like accessing any other Web page.
    \item Capable of running without the use of session cookies (page reload) or third party services to ensure patient's anonymity is upheld.
    \item Interaction with user-server should be one-off and response. Multiple interactions with the server are prevented by reducing the number of requests to the server.
\end{itemize}

The use of MERN (MongoDB, ExpressJS, ReactJS, NodeJS) stack enables a true separation of layers allowing flexible control over each front-end and back-end component as depicted in Fig. \ref{fig:CoughDetectApp}. \texttt{React} fundamentally uses SPA (Single Page Application) approach to quickly load a single resource (\texttt{index.js}) that contains the entire application rather than sending \texttt{HTTP} requests to the server every time a user wants to navigate elsewhere within the app. Not having to reload the page disables the need for storing cookies in the user machine that can be used to re-identify the user. The front-end logic is mainly \texttt{javascript} code that will be run in the client machine. 

Functional interaction, such as recording the cough, passing it over to the server for evaluation and receiving feedback, is done using a custom-built self-hosted API instance solution on a different port. Connections to the server are always encrypted using Hypertext Transfer Protocol Secure (\texttt{HTTPS}). Locally, in the server machine, the \texttt{node.js} endpoint interacts with a python-based API that implements the algorithms and methods from sections \ref{section:preprocessing} to \ref{section:classifier}. Once the server receives an audio stream, the processing pipeline is activated, a prediction of the test is issued, and an asynchronous message is returned to the user (client), through the same established secured connection, to update the Web App with the result of the test as illustrated in Fig. \ref{fig:CoughDetectApp}.
\section{Results and Evaluation}\label{section:results}
In this section we present A) the recognition results of DeepCough for the detection of Covid-19 vs non-Covid-19, and B) further categorisation of the  Covid-19 positive samples into groups indicating the grade of  Covid-19 disease, with respect to qRT-PCT and lymphocyte counts separately. A comparison of our proposed method DeepCough3D with approaches in related work (AI4COVID \cite{imran2020ai4covid}, Coswara \cite{sharma2020coswara}) and Cough against Covid \cite{CoughAgainstCovid_Bagad_2020}), as well as Auto-ML \cite{ASKL2} is also presented.
\subsection{Evaluation of Covid-19 Detection with DeepCough}
 The classification results are reported for a stratified $k=10$ cross-folding replication strategy for internal validity. A sample can only be exclusive member of one fold, (viz. folds are participant-independent). In each iteration a disjoint fold is left out for testing, a different one for validation and the remaining are used for training. The confusion matrix for DeepCough3D, shown in Table \ref{tab:confusionmatrix}, demonstrates the classification prowess of DeepCough3D with true positives at 97.18$\%$ and true negatives at 96.64$\%$.

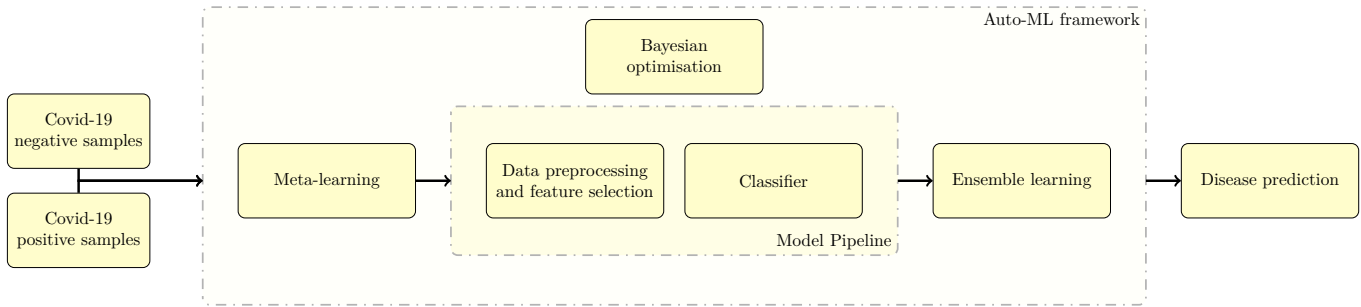
\begin{figure*}[!ht]
    \centering
    \resizebox{\textwidth}{!}{
    \begin{tikzpicture}
        \draw [dotted_block, fill = color1!10] (2.5,-3.5) rectangle (21.5,2.5) node[below left] {Auto-ML framework};
        \draw [dotted_block, fill = color1!50] (7.5,.5) rectangle (16.5,-2.5) node[above left] {Model Pipeline};

        \node [block, draw, rounded corners, text width = 75pt, text centered, fill = color1] (v0) at (0,-2) {Covid-19 positive samples};
        \node [block,draw, rounded corners, text width = 75pt, text centered, fill = color1] (v1) at (0,0) {Covid-19 negative samples};
        
        \node [block,draw, rounded corners, text width = 95pt, text centered, fill = color1] (v2) at (5,-1) {Meta-learning};
        
        \node [block,draw, rounded corners, text width = 95pt, text centered, fill = color1] (v3) at (10,-1) {Data preprocessing and feature selection};
        \node [block,draw, rounded corners, text width = 95pt, text centered, fill = color1] (v4) at (14,-1) {Classifier};
        \node [block,draw, rounded corners, text width = 95pt, text centered, fill = color1] (v4) at (12,1.5) {Bayesian optimisation};
        \node [block,draw, rounded corners, text width = 95pt, text centered, fill = color1] (v5) at (19,-1) {Ensemble learning};
        \node [block,draw, rounded corners, text width = 95pt, text centered, fill = color1] (v6) at (24,-1) {Disease prediction};
        
        \draw[myarrows] (v0) -- (0,-1) -- (2.5,-1);
        \draw[myarrows] (v1) -- (0,-1) -- (2.5,-1);
        \draw[myarrows] (21.5,-1) -- (v6);
        \draw[myarrows] (v2) -- (7.5,-1);
        \draw[myarrows] (16.5,-1) -- (v5);
        
    \end{tikzpicture}
    }
    \caption{Flowchart outlining the model selection process with Auto-ML \cite{ASKL2}.}
    \label{fig:auto_sklearn}
\end{figure*}

We also compare our proposed method with approaches in related work as well as AutoML \cite{ASKL2}. AutoML is a full model meta-learning algorithm that combines Bayesian optimisation in a set of shallow machine learning algorithms, such as k-nearest neighbours, na{\"i}ve Bayes, support vector machines, decision trees, random forest, and boosted classifiers. Auto-ML uses Bayesian optimisation of the AUC score to find for a method or their combinations (viz. pipelines), as well as the model hyper-parameters that yield the highest classification performance as delineated in Fig. \ref{fig:auto_sklearn}. It further considers feature selection through information gain, relief, $\chi^2$ statistics. The Auto-ML method is trained with flattened vectors of audio signal descriptors (Mel-Frequency Cepstral Coefficients, Zero-Crossing Rate, Roll-Off Frequency, and Spectral Centroid). 

Models were implemented in Python and trained on an Ubuntu Linux machine with AMD(R) Threadripper(R), 3.40GHz processor and 32 GB of RAM. Training time in this machine for 10-folds of the DeepCough approach was $\sim35$ minutes. We further deployed the trained models in an Oracle cloud virtual machine with eight cores (CPU only) as the back-end of the Web app (section \ref{section:weapp}. In this setting, the detection of a cough in a sound stream lasts in the range of 6-12 seconds and  the results of the test are issued in 1-2 seconds. 

\begin{figure*}[ht!]
    \centering
    \subfloat[][Comparison of DeepCough 3D and 2D Classifiers with an Auto-ML ensemble of shallow machine learning methods \cite{ASKL2} combined.]{
    \includegraphics[width=0.45\textwidth]{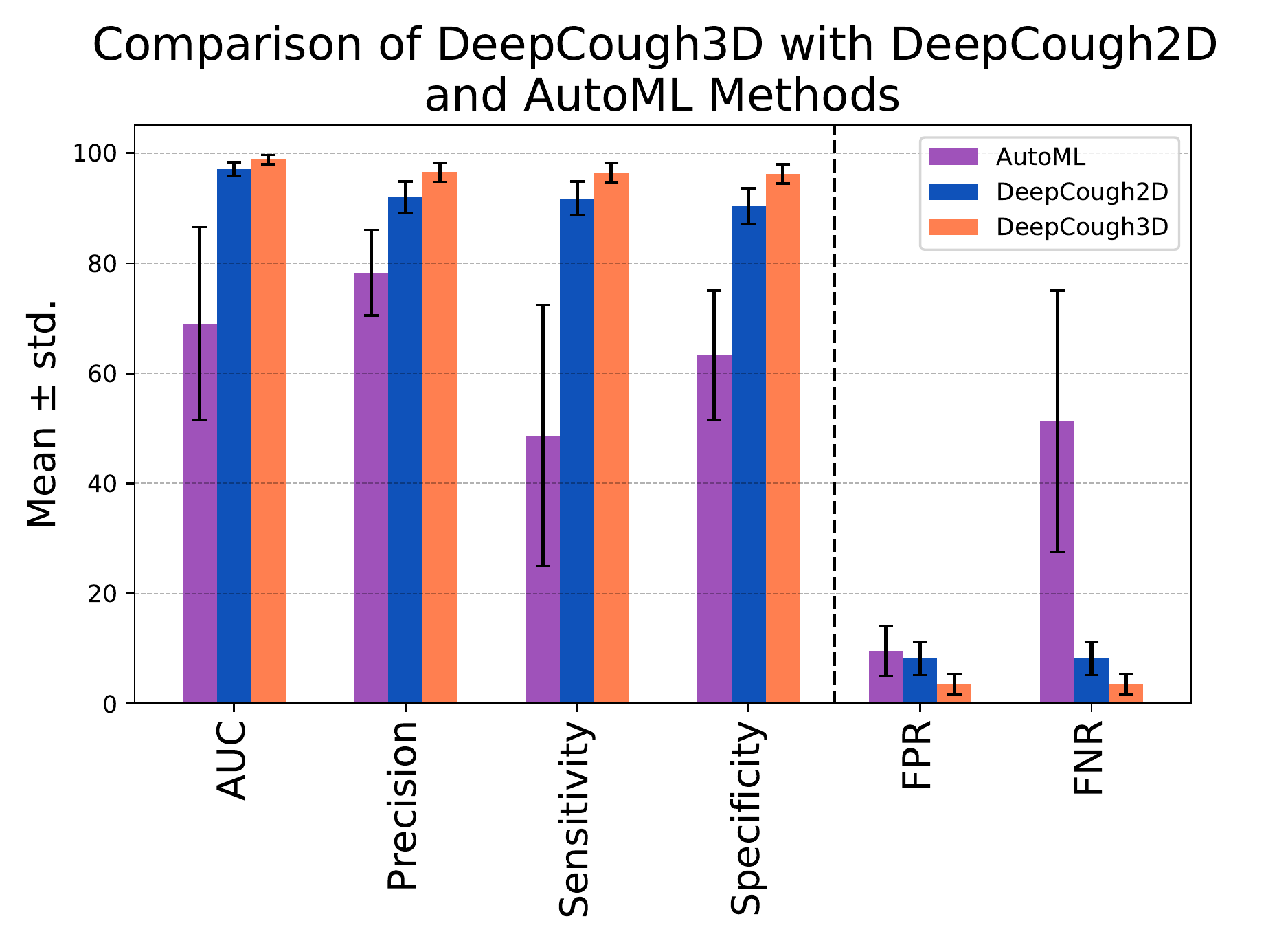}}
    \label{fig:ROCandbarplot}
    \hspace{1cm}
    \subfloat[][Receiver Operating Characteristic curve for DeepCough and other methods in related work as a function of true and false positive rates.]{
    \includegraphics[width=0.45\textwidth]{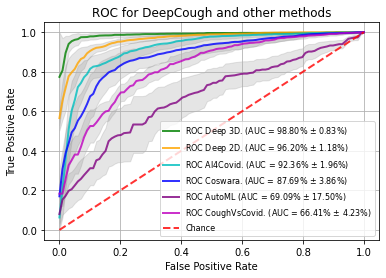}}
    \label{fig:seg}
\caption {\small a) Statistical metrics comparison of DeepCough with the other best and worst methods tested. b) Receiver Operating Characteristic (ROC) for DeepCough and other methods to detect pulmonary infection (Covid-19) coughs vs. other type of coughs using this study database.}
\label{fig:classificationdnn}
\end{figure*}

Performance comparison of DeepCough 2D and 3D vs other related approaches and Auto-ML in terms of statistical measures of AUC, precision, sensitivity, and specificity are listed in Table \ref{tab:Comparison_allmethods}.
A bar graph of the same results is shown in Fig. \ref{fig:classificationdnn} (a) for a visual comparison. In Fig. \ref{fig:classificationdnn} (b)
the recognition performance of DeepCough3D is primarily assessed in terms of the AUC since AUC allows considering both sensitivity and specificity for different cut-points and gives a better view of the benefit of the binary classifier with skewed samples, e.g. more negatives than positives, than standard accuracy. All of the above results conclude that DeepCough 3D approach afforded the highest significant performance rates in most statistics for the classification of Covid-19 positive vs negative cough samples.

\begin{table*}[!htbp]
\centering
\caption{\small A comparison of statistical performance measures of DeepCough3D with DeepCough2D,  AutoML \cite{ASKL2}, AI4COVID\cite{imran2020ai4covid}, Coswara \cite{sharma2020coswara}, and Cough Against (vs.) Covid \cite{CoughAgainstCovid_Bagad_2020} for recognition of Covid-19 coughs. The p-values for t-test are statistically significant (*: $p < 0.05$, **: $ p < 0.01$ ) for the average of all statistical metrics (M1-M4) for DeepCough3D in comparison to other methods. Likewise, the p-values for t-test for DeepCough2D are also reported.}
\renewcommand{\arraystretch}{1.1}
\scalebox{.8}{
\begin{tabular}{l l l l l |l l l l |l l l l }
& & & &&  \multicolumn{4}{c|}{DeepCough3D} & \multicolumn{4}{c}{DeepCough2D}\\
\textbf{ }& \textbf{AUC (M1)} & \textbf{Precision (M2)} &  \textbf{Sensitivity (M3)} &  \textbf{Specificity (M4)} & \textbf{M1} & \textbf{M2} &  \textbf{M3} &  \textbf{M4} & \textbf{M1} & \textbf{M2} &  \textbf{M3} &  \textbf{M4}\\ \hline
\textbf{DeepCough3D}    & 98.80 $\pm$ 0.83  & 96.54 $\pm$ 1.75      & 96.43 $\pm$ 1.85              & 96.20 $\pm$ 1.74 & - & - & -  &  - &** & **&** &** \\
\textbf{DeepCough2D}    & 96.20 $\pm$ 1.18  & 89.87 $\pm$ 1.46      & 89.63 $\pm$ 1.57              & 86.55 $\pm$ 4.64 & ** & ** &**  &**  &-  & - & - & -\\
\textbf{AutoML}         & 69.04 $\pm$ 17.50 & 78.28 $\pm$  7.78     & 48.70 $\pm$  23.71            & 63.26 $\pm$ 11.74 & ** & ** &**  &** & ** & * & ** & 0.22 \\
\textbf{AI4COVID}       & 92.36 $\pm$ 1.96  & 85.93 $\pm$ 2.87      & 85.87 $\pm$ 2.87              & 81.36 $\pm$ 4.31 & ** & ** &**  &**& ** & ** & ** & **\\ 
\textbf{Coswara}        & 87.69 $\pm$ 3.86  & 84.08 $\pm$ 3.57     &  81.99 $\pm$ 5.47             & 83.45 $\pm$ 3.48 & ** & ** &**  &** & 0.33& 0.14& 0.16&0.32\\ 
\textbf{Cough-vs-Covid} & 66.41 $\pm$ 4.23 & 76.04 $\pm$ 2.53       &  76.64 $\pm$ 2.28             & 67.00 $\pm$ 4.29 & ** & ** &**  &**& ** & ** & ** & **\\ \hline
\end{tabular}}
\label{tab:Comparison_allmethods}
\end{table*}

\begin{table}[!htbp]    
    \centering
    \scalebox{1.65}{ \hspace{-.5cm}
    \resizebox{5.1cm}{!} {
    \begin{tabular}{ l c c c }
    &\tiny \textbf{Positive} & \tiny \textbf{Negative} & \\
    \multirow{6}{*}{\textbf{\rotatebox[origin=b]{90}{\tiny Covid-19}}} &\cellcolor{blue!42} \tiny True Positive  & \cellcolor{blue!10} \tiny False Negative & \multirow{3}{*}{\tiny \textbf{Actual Positive}}\\
    &\cellcolor{blue!42} 97.18\%  & \cellcolor{blue!10} 2.82\% &\\
    &\cellcolor{blue!42} {\scriptsize 2273}  & \cellcolor{blue!10} {\scriptsize 66} &\\
    & \cellcolor{blue!12} \tiny False Positive & \cellcolor{blue!40} \tiny True Negative & \multirow{3}{*}{\tiny \textbf{Actual Negative}} \\
    &\cellcolor{blue!12} 3.36\% &\cellcolor{blue!40} 96.64\% &\\ 
    &\cellcolor{blue!12} {\scriptsize 203} & \cellcolor{blue!40} {\scriptsize 5838}& \\
    \end{tabular}}}
    \caption{\small Normalized confusion matrix for DeepCough3D.} 
    \label{tab:confusionmatrix}
\end{table}

\subsection{Assessing the grade of infection from Covid-19 positives samples} 
In this study, alongside the cough samples, we also collected the outcomes from \emph{quantitative real time polymerase chain test (qRT-PCR)} and \emph{lymphocyte count (blood ratio)} tests. qRT-PCR test is currently considered the gold standard for detecting a positive Covid-19 infection. qRT-PCR test detects the (Covid-19) virus' RNA within a patient's genetic material. In qRT-PCR test, the RNA is reverse transcribed to DNA using specific enzymes. Additional short fragments of DNA, that are complementary to transcribed viral DNA, are then added. Some DNA strands are programmed to release a fluorescent dye. The amount of fluorescence is monitored in each cycle, until a threshold is surpassed. The fewer the cycles (Ct) it takes to surpass this threshold, the higher the severity of the infection. During the Covid-19 pandemic, a challenge is to identify patients with low and mild levels of infection or asymptomatic, so called `carriers' \cite{day2020covid}. Regardless of their asymptomatic conditions, positive qRT-PCR detection can be done with an adequate sample pooling to deal with potential borderline Ct values from these patients \cite{lohse2020pooling}. For this experiment, we labelled a cough sample in terms of whether it came from a patient whose Ct values were borderline positive (30 $\leq$ Ct < 35), standard positive (20 $\leq$ Ct < 30) or strong positive (Ct $\leq$ 20). 

\begin{figure*}[!ht]
    \subfloat[][Sensitivity (recall), specificity and f1-score for the classification of positive cough samples labelled as borderline positive, positive and strong positives based on qRT-PCR results by DeepCough2D and DeepCough3D.]{%
        \includegraphics[width=0.25\textwidth, scale = 0.25, trim =.1cm .1cm .1cm .1cm, clip = true]{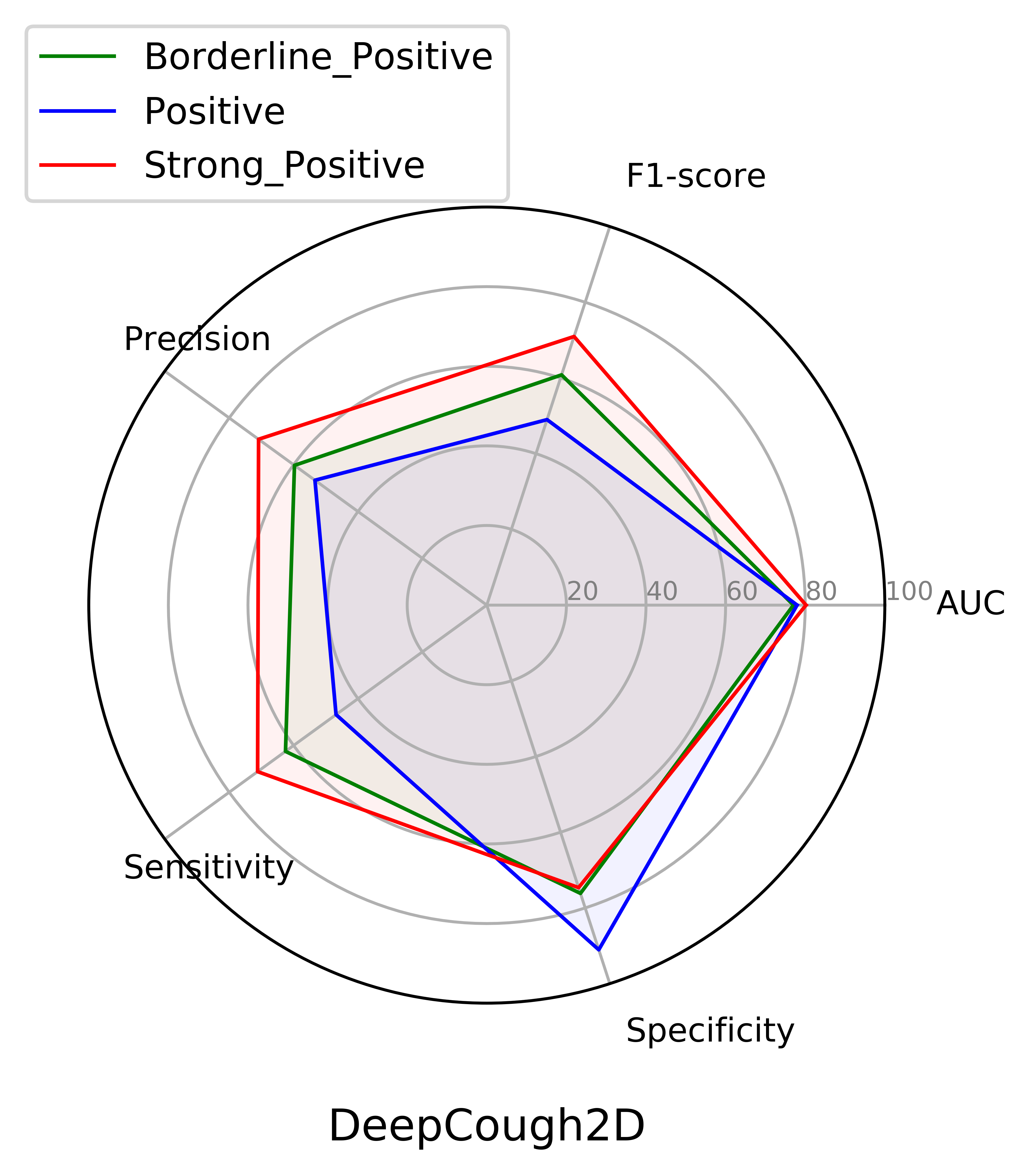}
        \includegraphics[width=0.25\textwidth, scale = 0.25, trim = .1cm .1cm .1cm .1cm, clip = true]{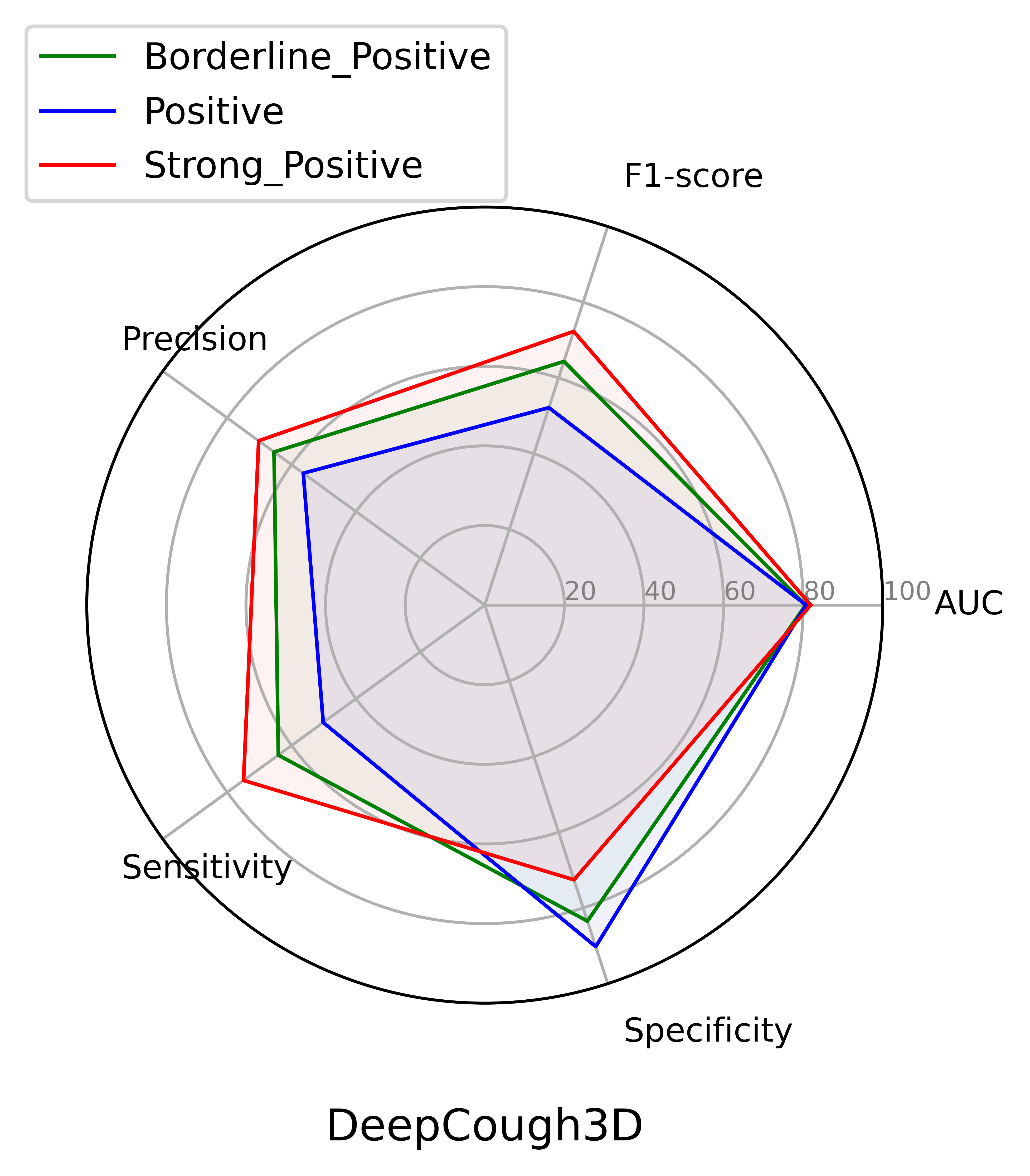}
        \label{fig:rtvalues}
    }
     \hspace{0.5cm}
    \subfloat[][Classification performance metrics for the recognition of Lymphopenia vs normal lymphocyte counts in blood sample.]{%
        \includegraphics[width=0.4\textwidth] {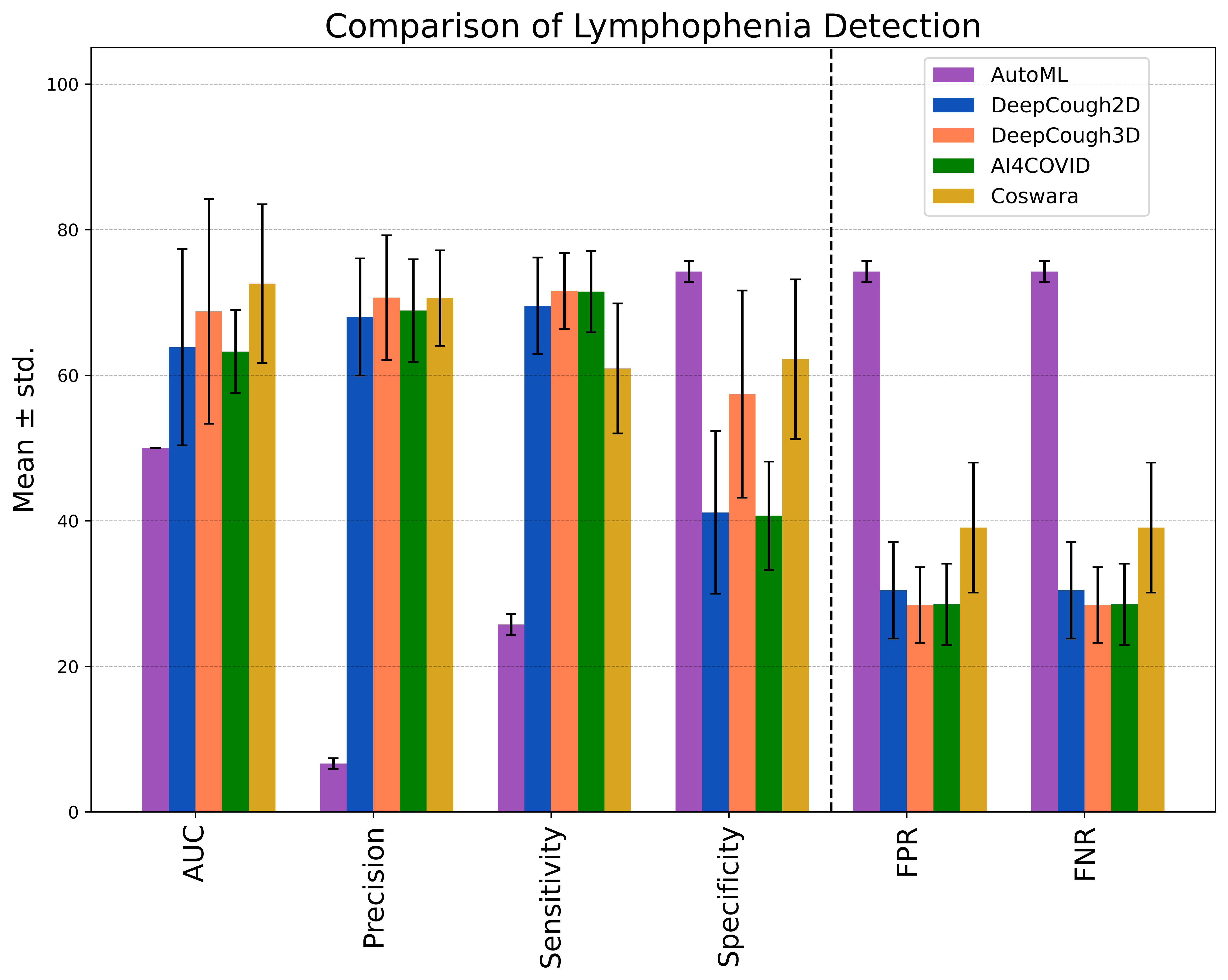}
        \label{fig:lymphocytes}
    }
    \caption{Statistical performance results for the recognition of possible markers of disease severity.}
\end{figure*}

The performance results for the recognition of the cough sample using DeepCough2D and DeepCough3D are displayed in Fig. \ref{fig:rtvalues} and enlisted in Table \ref{tab:RT_metrics}. Overall, performance results show a recognition rate well above chance level and average AUC of $81.08\%\pm 5.05\%$ for DeepCough3D. This can potentially be helpful to support two-stage screening protocols previously discussed in the introduction. The recognition of the disease severity was better at discriminating samples coming from the highly discerning groups, i.e. borderline and high positive, but struggled with the intermediate group. This is, however expected as intermediate samples can have a mixed pattern of cough acoustics to those loosely or highly affected. Nevertheless, its specificity was highly better than for the two other groups.

\begin{table}[!htbp]
    \hspace{-1cm}
    \caption{\small Statistical metrics for the classification results of positive cough samples labelled as borderline positive, standard positive and strong positives based on qRT-PCR results by DeepCough2D and DeepCough3D.}
    \label{tab:RT_metrics}
    \renewcommand{\arraystretch}{1.75}
    \scalebox{0.68}{
    \begin{tabular}{l l l l l l}
        \textbf{DeepCough} & \textbf{Statistical} & \textbf{Borderline}               & \textbf{Standard}                      & \textbf{High}    & \textbf{Average}  \\
        \textbf{Model} & \textbf{Metric}                       & \textbf{Positive}                 & \textbf{Positive}                      & \textbf{Positive}  & \textbf{of 3 classes}\\ \hline
        \multirow{5}{*}{DeepCough2D}&             \textbf{AUC} &         \( 77.05\pm3.70\%\)  &      \( 77.95\pm7.77\%\)  &    \(80.2\pm3.78\%\) & \(78.4\pm5.08\%\)\\
         & \textbf{F1-score} &         \(60.81\pm7.48\%\)  &     \( 48.98\pm11.29\%\)  &  \( 70.94\pm4.41 \%\) &\(60.24\pm7.73\%\)\\
        & \textbf{Precision} &         \( 59.72\pm6.90\%\)  &     \( 53.36\pm11.05\%\)  &  \( 70.88\pm4.51\%\) &\(62.32\pm7.49\%\) \\
      & \textbf{Sensitivity} &         \(62.51\pm9.78\% \) &     \( 46.81\pm13.42\% \) &  \(  71.17\pm5.40\%\) & \(60.83\pm9.53\%\)  \\
      & \textbf{Specificity} &         \(76.15\pm5.17\%\)  &     \(  91.04\pm3.74\% \) &  \( 74.63\pm4.87\% \) &\(80.61\pm4.59\%\) \\
        \hline
         \multirow{5}{*}{DeepCough3D}&\textbf{AUC}   &        \( 80.59\pm5.71\%\) &      \( 80.75\pm5.34 \%\)&   \(81.09\pm4.11\%\)& \(81.08\pm5.05\%\) \\
          & \textbf{F1-score} &        \(64.37\pm10.89\% \)&      \( 52.15\pm8.52\% \)&    \(72.3\pm5.76\%\) & \(62.94\pm8.39\%\) \\
         & \textbf{Precision} &        \(65.44\pm10.17\%\) &     \(  56.37\pm8.31\%\) &  \( 70.23\pm6.35\%\) &\(64.01\pm8.28\%\) \\
       & \textbf{Sensitivity} &        \(64.12\pm13.21\% \)&      \(50.18\pm12.35\%\) &  \( 74.93\pm7.22\% \)&\(63.08\pm10.93\%\) \\
       & \textbf{Specificity} &        \( 83.47\pm4.04\% \)&      \( 90.22\pm3.13\%\) &  \( 72.56\pm5.28\%\)&\(82.08\pm4.15\%\) \\          \hline
    \end{tabular}}
\end{table}

Another marker of disease severity that we have explored is lymphocyte count (viz. lymphopenia vs normal levels of lymphocyte counts). Lymphopenia is a condition defined as when patients have a blood lymphocyte percentage (LYM\%) lower than 20\%. Lymphopenia is the frequency associated with a severe infection or illness. The performance results from the recognition of lymphopenia vs normal levels of lymphocytes are graphically displayed in Fig. \ref{fig:lymphocytes}. Although some works have suggested lymphocyte count as a way to grade Covid-19 severity \cite{huang2020lymphopenia}, our results to predict an infection grade using this marker are not as good as when labelled by the qRT-PCR test. However, the performance of DeepCough3D could also be hampered, at this occasion, by subset levels of lymphocyte counts that can be affected by biological and inter-subject variabilities \cite{tosato2013biological, lymphocytopeniaNIH}.

\section{Discussion}
\label{section:discussion}
The Covid-19 pandemic has proven difficult to contain not only because of its high infection rate, but also because the symptoms of Covid-19 borne stark similarities with other viruses such as the common-flu and pneumonia. Hence, it has been particularly challenging for carriers of Covid-19 to know that they have been infected by Covid-19, therefore furthering the spread of Covid-19. To facilitate the early detection of Covid-19, we have developed a test from clinically validated Covid-19 positive and negative individuals that provided a cough sample and performed a molecular-based test in certified laboratories. 

 This is a multi-center study, with populations from Spain and Mexico, to ensure the trained inference mechanism of DeepCough3D is unbiased towards particular demographic characteristics. In addition, the proposed DeepCough3D model, subsequently embedded in CoughDetect for recognition of Covid-19 coughs, was compared against related work and Auto-ML \cite{ASKL2}. AutoML is a method for algorithm selection and hyper-parameter tuning, optimised through a full model selection strategy. In all the performance metrics for Covid-19 positives recognition, DeepCough3D performed better, as noted in Table \ref{tab:Comparison_allmethods}. The reported results reaffirm that DeepCough3D learning method used is successful in distinguishing between Covid-19 positive and negative cough samples. 

The performance of DeepCough3D for establishing whether a given cough sample is from a Covid-19 positive or negative patient is clinically sound as a \emph{primary test or pre-screening} with an AUC of 98.80 $\pm$ 0.83, a sensitivity of $96.43\% \pm 1.85\%$, and a specificity of $96.20 \% \pm 1.74\%$. The strength of DeepCough3D lies in high recognition performance over a large set of clinically validated cough samples earmarked with molecular test. This resolutely corroborates the informational potential of the latent audio sonographs of coughs to detect an acute pulmonary disease such as Covid-19. The diagnostic sensitivity of the gold-standard molecular test for Covid-19, i.e. qRT-PCR, is $98\%$ for nasopharyngeal swab tests, whereas for saliva is $91\%$ \cite{euRTPCR}. However, the reported averaged sensitivity of commercial serological kits (e.g. based on lateral flow immunoassays) for Covid-19 was only $65\%$ average ($49.0\%$ min. to $78.2\%$ max.) \cite{bastos2020diagnostic}.
\section{Conclusion}
\label{section:conclusion}
In this work, a primary screening test for Covid-19 is proposed and assessed using clinically validated cough samples of participants, who jointly performed a molecular-test (qRT-PCR) in our partner hospitals. The proposed test framework is powered by a generic cough identification algorithm based in EMD and a recognition method named DeepCough3D. This latter method generates a 3D audio tensor to leverage the strength of a convolutional neural network approach to identify the latent characteristics in Covid-19 cough signals. The performance of DeepCough3D attains an AUC of 98.80 $\pm$ 0.83, a sensitivity of $96.43\% \pm 1.85\%$, that is comparable to the reported sensitivity ($91\%\pm10\%$) of accelerated serology tests based on saliva \cite{czumbel20}. The proposed generic method does not require using specific transfer learning models or data from other studies, paving the way for derivative works. The proposed approach outperforms related works and other state-of-the-art methods. Further, the quality of our clinically controlled and validated large dataset increases our confidence in the validity of these results.

In addition to the development of a recognition test for Covid-19 using coughs, this work further investigates the possibility to recognise the extent of the Covid-19 infection in Covid-19 positive participants. This is undertaken with the qRT-PCR test and the lymphocyte count, and the results greatly surpassed chance levels of performance, indicating the feasibility of assessing severity to some extend. Classification of the coughs in three severity levels, defined by the resulting Ct of the molecular test for Covid-19, yields an average AUC of $81.08\%\pm 5.05\%$. This could potentially serve as an additional functional feature to diagnose the extent of the Covid-19 infection in a given Covid-19 carrier. This can help facilitate effective management of healthcare facilities during a pandemic, such as ventilators, which were in short supply during the first wave of the Covid-19 pandemic around the world.

Furthermore, the entire framework has been embodied as a web-app service available at \href{http://coughdetect.com}{CoughDetect.com}. The motivation for developing this alternative test based on coughs is to have a fast turn-around for Covid-19 point-of-need primary test to a) reduce the burden on specialised personnel for clinical or secondary diagnosis of Covid-19, b) to make the primary screening available to masses at large from the comfort of their homes at negligible costs, and c) the anonymity of the participants is kept at the core by using in-house custom code to power the analysis and recording only their cough sounds. It can also be used as an electronic health certificate at public places such as airports, and schools.

In the midst of a global pandemic, the significance of our proposed point-of-need primary test, developed and tested on clinically validated data, is paramount. Our proposed primary test can mitigate the logistics, long turn-around time, and cost of clinical diagnostic test of Covid-19. For future works, parameter tuning of the sonograph representations and complementary analysis of coughing behaviours could be explored to investigate further improvements in performance. It would also be of interest to investigate whether tracking of Covid-19 progression can be done using DeepCough3D. 

\textbf{Acknowledgements. } We thank the reviewers for the helpful and constructive comments. Thanks are extended  to the ESRC Impact Acceleration Account (ES/T501815/1) at Essex University and the Talentia Senior Program by The Regional Ministry of Economy, Innovation, Science and Employment of Andalusia (reg. 201899905497765) for indirectly supporting this research. Thank you to Oracle for Research for providing Oracle Cloud credits and related resources to support this project. 

\bibliographystyle{IEEEtran}
\bibliography{IEEEabrv, CoughDetect}
\vspace*{-0.8em} 

\appendix

\section*{Author contributions}  
\small{ \emph{JAP} contributed to the conceptualisation and coordination of the work, methodology, implementation, analysis, figures and writing of the manuscript. \emph{HPE} contributed to the organisation of the study, methodology, implementation, analysis, figures and writing of the manuscript. \emph{ET}, \emph{MGP} \& \emph{ABT} worked in the data collection and laboratory analysis. \emph{MK} was involved in the writing of the manuscript, elaboration of figures and analyses. \emph{ARP, ORG,} \& \emph{ATG} contributed to the implementation of the proposed approach, and comparison methods. \emph{DJ} contributed to the signal processing and cough sound detection. \emph{ZA} worked in the sonograph analysis. \emph{NG} worked in the implementation of the web-app prototype. \emph{CGR} contributed to the coordination, methodology and appraisal of the work.}

\renewcommand{\arraystretch}{1.5}
\begin{table}[h]
    \centering
    \begin{tabular}{wl{1.5cm} p{6.8cm}}
         \raisebox{-.9\height}{\includegraphics[width=0.7in,clip,keepaspectratio]{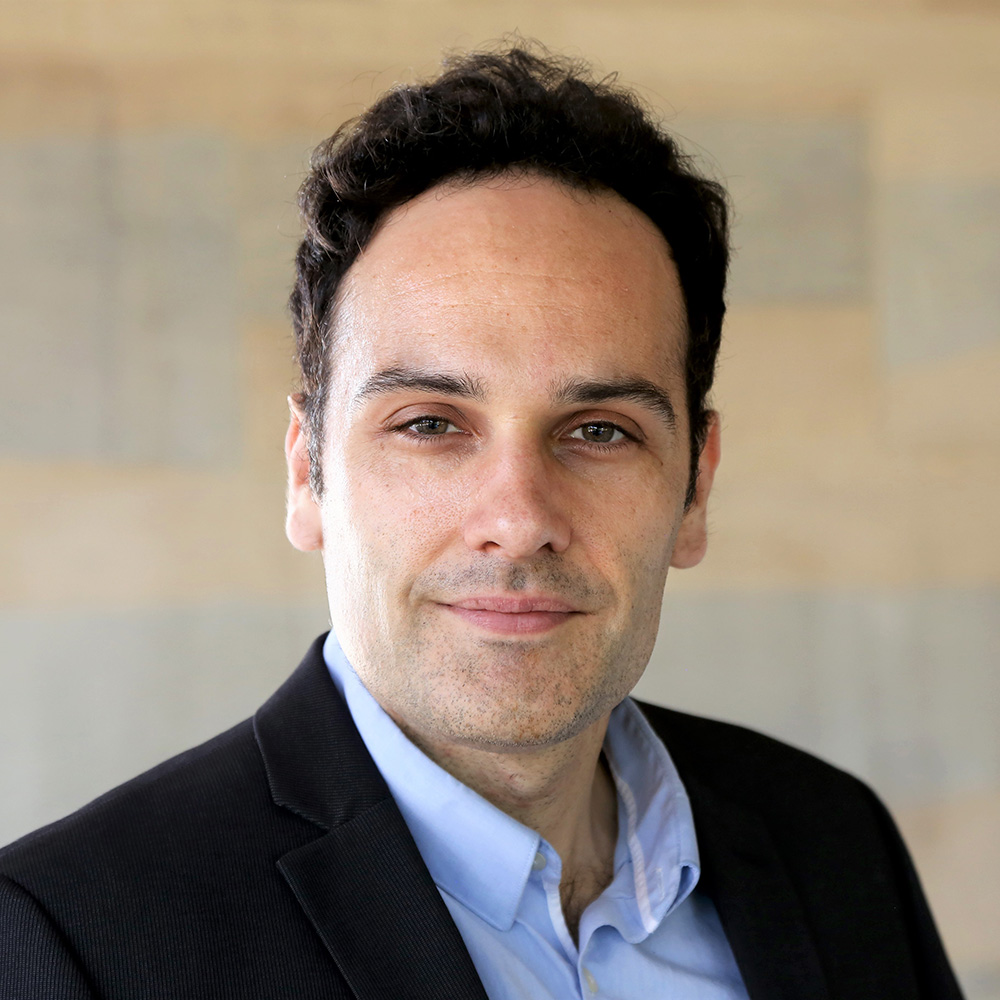}}  &
         \small{{\bf Javier Andreu-Perez} (\emph{PhD, SMEEE '19, HEA '19, ACM EPSRC '20}) is Senior Lecturer (Assoc. Prof.) at University of Essex (UK), chair of the Smart Health Technologies Group and Talentia Fellow at Simbad2, University of Jaén (Spain)}  
         \\
         \raisebox{-.9\height}{\includegraphics[width=0.7in,clip,keepaspectratio]{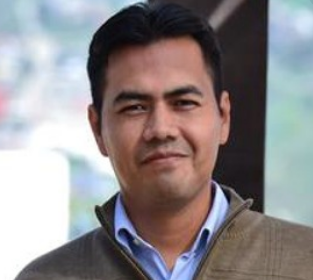}}  &
         \small{{\bf Humberto Perez-Espinosa} (\emph{PhD}) is Researcher at CICESE-UT3 (Nayarit, Mexico)  in intelligent audio analysis. He is member of the Mexican National System of Researchers and Visiting Fellow at University of Essex (UK)‬}
         \\
         \raisebox{-.9\height}{\includegraphics[width=0.7in,clip,keepaspectratio]{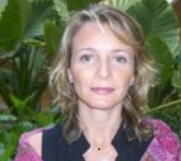}}  &
         \small{{\bf Eva Timonet} (\emph{PhD, RN}) is Head of the Nursing Unit at the Junta de Andalucia, Health Agency Costa del Sol and Senior Research Scientist at the Institute of Biomedical Research in Malaga (IBIMA) (Spain).‬}
         \\
         \raisebox{-.9\height}{\includegraphics[width=0.7in,clip,keepaspectratio]{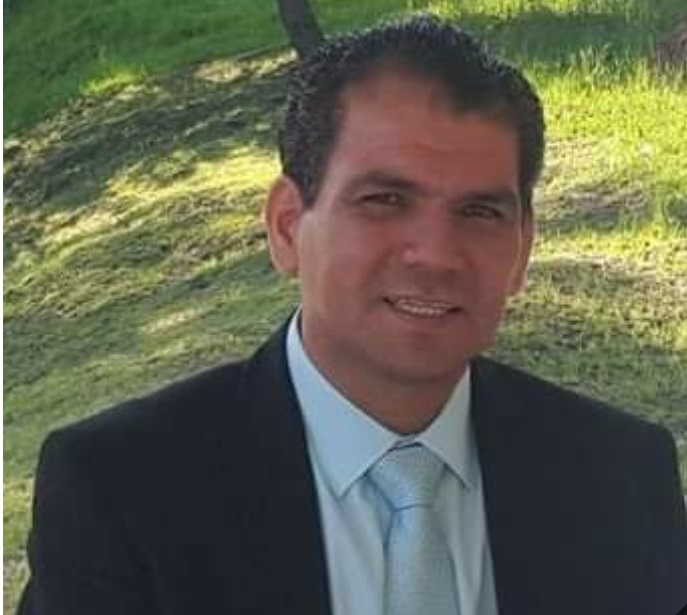}}  &
         \small{{\bf Manuel I. Giron-Perez} (\emph{PhD}) is Clinical Professor of the Autonomous University of Nayarit. He is member of the Mexican National Academy of Medicine, the Mexican Academy of Science, and National System of Researchers.‬}
         \\
         \raisebox{-.9\height}{\includegraphics[width=0.7in,clip,keepaspectratio]{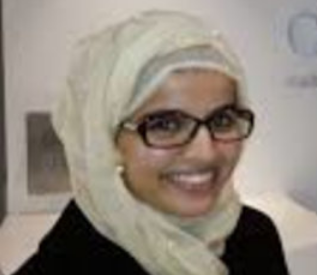}}  &
         \small{{\bf Mehrin Kiani} (\emph{MSc, PhD std.}) is PhD student in computational intelligence for health sciences at the The Smart Health Technologies Group at University of Essex (UK).‬}
         \\
         \raisebox{-.9\height}{\includegraphics[width=0.7in,clip,keepaspectratio]{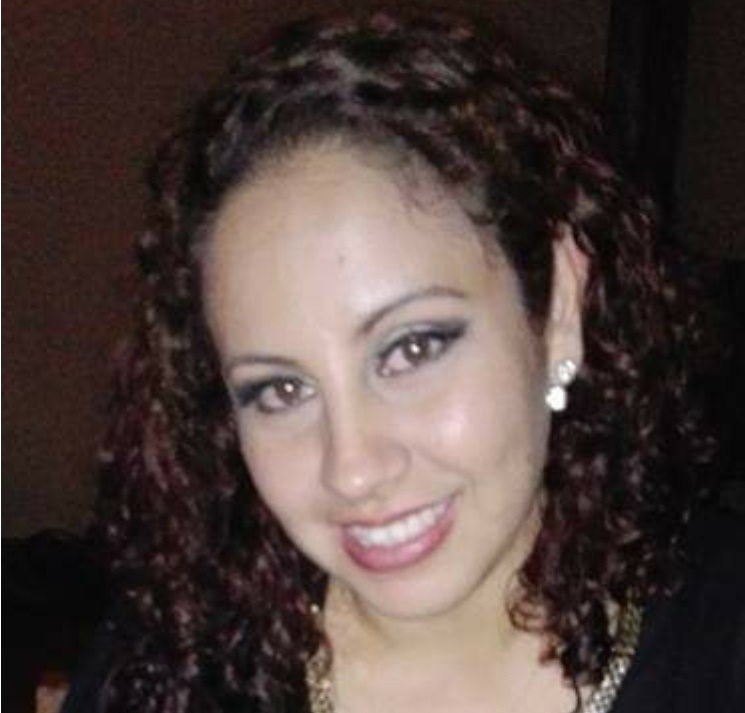}}  &
         \small{{\bf Alma B. Benitez-Trinidad } (\emph{PhD}) is  Research associate in Toxicology. She is a member of National System of Researchers of Mexico Mexico and professor of the Autonomous University of Nayarit.‬}
         \\
         \raisebox{-.9\height}{\includegraphics[width=0.7in,clip,keepaspectratio]{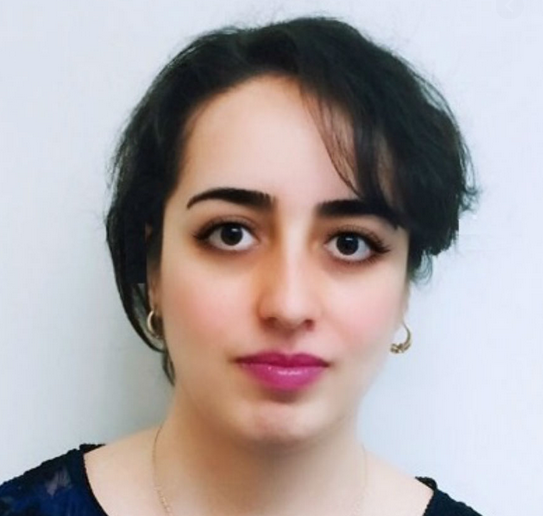}}  &
         \small{{\bf Delaram Jarchi} (\emph{PhD, SMIEEE'18}) is Lecturer (Assistant Professor) in Advanced Signal Processing at University of Essex (UK) and member of the Embedded Technologies and Smart Health Technologies Group.‬}
         \\
         \raisebox{-.9\height}{\includegraphics[width=0.7in,clip,keepaspectratio]{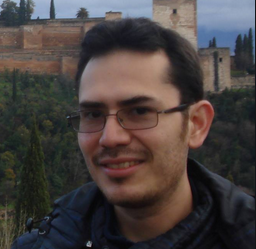}}  &
         \small{{\bf Alejandro Rosales-Perez} (\emph{PhD}) is Senior Research Scientist at CIMAT (Monterrey, Mexico) in Machine Learning and Pattern Recognition. He is member of the Mexican National System of Researchers.‬}
         \\
         \raisebox{-.9\height}{\includegraphics[width=0.7in,clip,keepaspectratio]{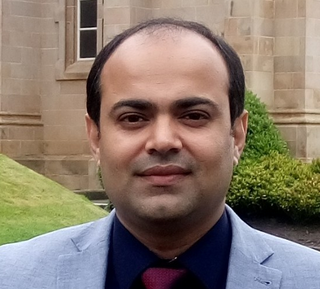}}  &
         \small{{\bf Zulfiqar Ali} (\emph{PhD}) is Lecturer (Assistant Professor) at University of Essex (UK), in Digital Speech Processing, Privacy Protection and Audio Forensics.‬}
         \\
         \raisebox{-.9\height}{\includegraphics[width=0.7in,clip,keepaspectratio]{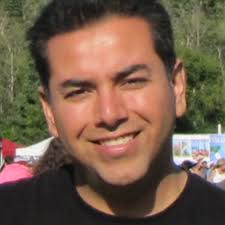}}  &
         \small{{\bf Orion Reyes-Galaviz} (\emph{PhD}) is Senior Machine Learning Engineer at Laivly (Canada), obtained his PhD from University of Alberta (Canada), where he was research assistant, and he is associate investigator at INAOE.‬}
         \\
         \raisebox{-.9\height}{\includegraphics[width=0.7in,clip,keepaspectratio]{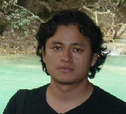}}  &
         \small{{\bf Alejandro Torres-Garcia} (\emph{PhD}) is research assistant in Biosignal Processing at INAOE where he recieved his PhD in Computational Intelligence and associate researcher at NTNU (Norway)‬}
         \\
         \raisebox{-.9\height}{\includegraphics[width=0.7in,clip,keepaspectratio]{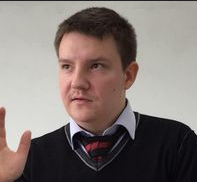}}  &
         \small{{\bf Nick Gatzoulis} (\emph{BSc, MSc std.}) is research officer at University of Essex (UK) and full stack software developer in web applications applications.‬}
         \\
          \raisebox{-.9\height}{\includegraphics[width=0.7in,clip,keepaspectratio]{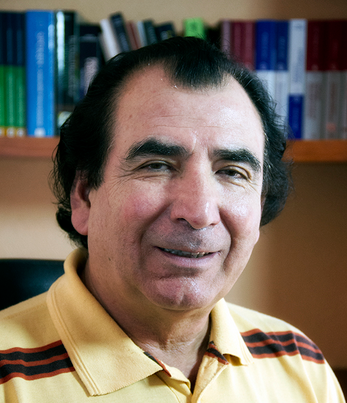}}  &
         \small{{\bf Carlos Reyes-Garcia} (\emph{PhD}) is Professor and Senior Investigator in Biosignal Processing at INAOE. He obtained his PhD from Florida State University (USA) in Computational Intelligence. His research interests are intelligent spectrographic audio analysis.‬}
    \end{tabular}

\end{table}

\end{document}